\documentclass[twocolumn]{aastex631}

\newcommand{\rpeak}{$R^{{\mathrm{X}}}_{\mathrm{peak}}$}

\definecolor{navyblue}{rgb}{0.0, 0.0, 0.5}

\newcommand{\magarc}{mag~arcsec\ensuremath{^{\mathrm{-2}}}}

\usepackage{amsmath}

\makeatletter
\DeclareRobustCommand{\HI}{%
  \mbox{H\check@mathfonts\fontsize\sf@size\z@\selectfont I}%
}
\makeatother

\begin{document}

\title{Truncations in the X-ray Halos of Early-Type Galaxies as a Tracer of Feedback and Mergers}

\author[0000-0002-1598-5995]{Nushkia Chamba}
\thanks{NASA Postdoctoral Program Fellow}
\affiliation{NASA Ames Research Center,  Space Science and Astrobiology Division M.S. 245-6, Moffett Field, CA 94035, USA}
\correspondingauthor{Nushkia Chamba}
\email{nushkia.chamba@nasa.gov}

\author{Pamela M. Marcum}
\affiliation{NASA Ames Research Center,  Space Science and Astrobiology Division M.S. 245-6, Moffett Field, CA 94035, USA}

\author[0000-0003-3249-4431]{Alejandro S. Borlaff} 
\affiliation{NASA Ames Research Center, Space Science and Astrobiology Division M.S. 245-6, Moffett Field, CA 94035, USA}
\affiliation{Bay Area Environmental Research Institute, Moffett Field, California 94035, USA}

\author[0000-0002-8341-342X]{Pasquale Temi}
\affiliation{NASA Ames Research Center,  Space Science and Astrobiology Division M.S. 245-6, Moffett Field, CA 94035, USA}

\author[0000-0002-0905-7375]{Aneta Siemiginowska}
\affiliation{Harvard Smithsonian Center for Astrophysics, 60 Garden St, Cambridge, MA 02138, USA}





\begin{abstract}


The morphology of X-ray halos in early-type galaxies depends on key structure assembly processes such as feedback and mergers. However, the signatures of these processes are difficult to characterize due to their faint and amorphous nature. We demonstrate that the truncation in the temperature profile of X-ray halos, defined by the radial location of the peak temperature, is significantly more impacted by recent mergers or galaxy interactions than feedback processes. At a fixed stellar mass, a highly asymmetric X-ray halo can be nearly a factor of ten more truncated than a relaxed one. This analysis led to a discovery of previously unknown asymmetric features in the optical and X-ray halos of three massive galaxies. We detect the intra-group star light and a large $\sim$45\,kpc size stellar stream connected to NGC~0383, suggesting that a recent stellar accretion event has triggered its active galactic nuclei to emit a powerful radio jet. While the disturbed X-ray halo of NGC~1600 is also related to a galaxy-satellite tidal interaction detected in optical imaging, the X-ray shape and asymmetry of NGC~4555 is highly unusual for a galaxy in a low dense environment, requiring further investigation. These results highlight the importance of truncations and deep imaging techniques for untangling the formation of X-ray halos in massive galaxies.

\end{abstract}




\section{Introduction} \label{sec:intro}

The assembly of hot $T\sim 10^7$\,K X-ray halos in early-type galaxies (ETGs) depend on key processes such as feedback \citep[due to star formation or active galactic nuclei (AGN);][]{bower+2006mnras370_645, diehl+2008apj680_897}, gas accretion into black holes \citep{silk+1998aap331_1, croton+2006mnras365_11}, mergers \citep{lacey+1993mnras262_627, sinha+2009mnras397_190}  as well as dark matter content \citep[e.g.][]{buote+2012inbook_235}. Consequently, the morphology of X-ray halos in surface brightness or temperature are not always smooth or relaxed, with asymmetric arcs and tails \citep[e.g.][]{trinchieri2004apss289_367} which can be very faint $\sim 10^{-8}-10^{-11}$ s$^{-1}$ cm$^{-2}$ arcsec$^{-2}$ \citep{borlaff+2024apj967_169}. Likewise, the shells, tails or the faint glow of intra-group/cluster light seen in optical images where the surface brightness $\mu_g \lesssim 26.5$~\magarc\ are signs of different stages of merging \citep[see][]{duc+2015mnras446_120, spavone+2017aap603_38} or the debris of large-scale dark matter and stellar structure assembly \citep[][]{montes+2019mnras482_2838, alonsoasensio+2020mnras494_1859, butler+2025arXiv2504.03518}. \par  \par

Despite these asymmetries, the temperature of X-ray halos in ETGs is thought to follow a ``universal'' 1D radial profile shape. This profile features an inner dip ($R^X_{\rm break}$)  and a peak (\rpeak) at a larger radius, with average values of 4 and 35\,kpc, respectively, and  peak temperature $<T^X_{\rm peak}> \sim 1.3-1.4$\,keV \citep[][]{kim+2020mnras492_2095}. While rising temperatures within $R^X_{\rm break}$ can be related to recent star formation, no clear evidence for the impact of AGN feedback or mergers on the temperature profile has been reported \citep[see][]{kim+2020mnras492_2095}.

This paper is motivated by the hypothesis that \rpeak{}, the radial location of the X-ray halo peak temperature, is governed by external encounters such as mergers and galaxy interactions or environmental processes such as ram pressure \citep[see][]{kim+2020mnras492_2095}. As described above, these processes can also lead to large-scale asymmetries in the X-ray halo \citep[see e.g.][]{kolokythas+2020mnras496_1471, islam+2021apj256_22, brienza+2022aap661_92}. Within this context, we investigate how the location of the \rpeak{} radius is linked with X-ray halo asymmetry.\par 
Parametric measures of galaxy asymmetry has a long history in optical astronomy, starting from the works by for e.g. \citet{conselice+2000apj529_886} and have since been applied to radio frequencies using maps of the neutral atomic hydrogen component \citep{holwerda+2011mnras416_2401}. The parameters used in those studies such as rotational asymmetry have been well tested using simulations at these wavelengths and many more asymmetric measures have since been proposed \citep[e.g.][and references therein]{pawlik+2016mnras456_3032, rodriguezgomez+2016mnras458_2371, sazonova+2024afz7_77}. The methods presented here are one of the first attempts to measure the asymmetry of galaxies in the X-ray domain.

Using archival observations from the \emph{Chandra} X-ray Observatory (Sect. \ref{sect:data}), we demonstrate that ETGs with significant asymmetry in the outer regions of their X-ray halos ($A_S > 0.4$) are those with very small \rpeak{} $<< 35\,$kpc. In other words, their peak X-ray temperature radius is much smaller than the average value from the universal temperature profile \citep{kim+2020mnras492_2095}, indicating that asymmetric X-ray halos are more truncated (Sect. \ref{sect:results}). \par 
We additionally present new asymmetric features in the truncated X-ray halos of three ETGs, NGC~0383, 1600 and 4555 (Sect. \ref{sect:results}). We discuss how these features can be linked with recent galaxy interactions or ongoing mergers (Sect. \ref{sect:discussion}).  While measuring asymmetry requires intensive image processing techniques that preserve low surface brightness (LSB) emission in images \citep{borlaff+2024apj967_169}, \rpeak{} offers a more scalable and easy-to-use metric for merger history that could persist long after more conventional signatures in the stellar and neutral plasma components have faded below detection.

\section{Data and Methodology}
\label{sect:data}

To study the distribution of the X-ray halo peak temperature radius (\rpeak{}), we utilize (1) the stellar size--stellar mass relation ($R^S_{\rm edge}-M_{\star}$; Sect. \ref{sect:size_plane}) and (2) X-ray halo asymmetry (Sect. \ref{sect:saunas}). The $R^S_{\rm edge}-M_{\star}$ relation describes galaxy growth in the outskirts \citep[e.g.][]{whitney+2019apj887_113}  due to mergers \citep{buitrago+2017mnras466_4888, trujillo+2020mnras493_87} and feedback processes \citep{chamba+2024apj974_247}. \par 
The sensitivity of the distribution of galaxies in the $R^S_{\rm edge}-M_{\star}$ plane to these processes which can also impact X-ray halos make the relation ideal for this study. 
Studying the distribution of \rpeak{} with respective to the $R^S_{\rm edge}-M_{\star}$ plane also provides insight as to where the X-ray halo peak temperature radius (\rpeak{}) is spatially located (either within or outside the stellar boundaries of ETGs). The distribution of \rpeak{} is specifically examined with respect to the ETG stellar size--mass plane. \par 

Given that feedback and mergers can also lead to disturbances or asymmetries in the X-ray halo (see Introduction), the dependence of \rpeak{} on the total luminosity of the hot plasma \citep[$L_X$ taken from][]{kim+2019apj241_36} and on the asymmetry of the X-ray halo shape ($A_S$) are also considered. While galaxy shape asymmetry has traditionally been applied to optical imaging \citep[see e.g.][]{pawlik+2016mnras456_3032, rodriguezgomez+2019mnras483_4140, sazonova+2024afz7_77},  here we devise a simple strategy to measure X-ray halo shape asymmetry using surface brightness maps. $L_X$ and $A_S$ are then used to interpret the location and scatter of \rpeak{} in the $R^S_{\rm edge}-M_{\star}$ plane. The data and methods used for the above tasks are described below in Sect. \ref{sect:size_plane} and \ref{sect:saunas}.

\subsection{\rpeak{} and the stellar size--mass plane}
\label{sect:size_plane}

\rpeak{} radii measurements are taken from \citet[][labeled as ``$R_{\rm MAX}$'' in that work]{kim+2020mnras492_2095}. The full \citet{kim+2020mnras492_2095} sample consists of 60 ETGs located in low mass group and cluster environments for which temperature profiles were created as part of the Chandra Galaxy Atlas\footnote{See  \protect\url{https://cxc.cfa.harvard.edu/GalaxyAtlas/v1/cga_main.html}} \citep[hereafter, full CGA sample;][]{kim+2019apj241_36}. Using these profiles, \rpeak{} has only been clearly identified in 30 ETGs by \citet{kim+2020mnras492_2095} (hereafter, the \rpeak{} sample). Table 1 in \citet{kim+2020mnras492_2095} list other X-ray properties of the sample such as the total luminosity ($L_X$) of the hot plasma. Estimates of the total mass are not available for the full sample. \par 
Stellar mass estimates of these 30 galaxies are taken from the $z = 0$ Multiwavelength Galaxy Synthesis catalogue  \citep{leroy+2019apj244_24}, the Spitzer Survey of Stellar Structure in Galaxies \citep[S4G;][]{sheth+2010pasp122_1397, munozmateos+2013apj771_59, querejeta+2015apj219_5, watkins+2022aap660_69} and from the compilation in the Chandra Nearby Galaxies catalogue \citep{bi+2020apj900_124} if not available in the former. The stellar masses of this sample span the range $10^{10.5} - 10^{12}\,M_{\odot}$. \par

To represent the outer stellar sizes of galaxies, the size--stellar mass plane is defined using the galaxy ``edge'' relations published in \citet{trujillo+2020mnras493_87, chamba+2022aap667_87} for field ETG and late-type galaxies (LTG). Briefly, the concept of stellar ``edge'' is physically motivated by theoretical \citep{schaye2004apj609_667} and observational  \citep{martinezlombilla+2019mnras483_664, diazgarcia+2022aap667_109} evidence for the star formation threshold. Following these works, the location of the truncation \citep{vanderkruit1979aap38_15} is used as a proxy for the location of the edge size definition ($R^S_{\rm edge}$) and can be observed as a change in slope in the radial color profiles of galaxies. \par
For the purpose of verifying that these edge scaling relations are also applicable to CGA galaxies, $R^S_{\rm edge}$ is measured using publicly available multi-wavelength images in the Spitzer IRAC 3.6$\mu$m and 4.5$\mu$m bands from the ETG Spitzer Survey of Stellar Structure in Galaxies \citep[18 galaxies in the full CGA sample and  six in the \rpeak{} sample;][]{watkins+2022aap660_69} and  g- and r-band imaging from the Dark Energy Camera Legacy Survey \citep[DECaLS; 24 galaxies in the \rpeak{} sample;][]{dey+2019aj157_168}. DECaLS can also be used detect stellar tidal features in ETGs which are brighter than $\mu_r > 28.4$~\magarc\ ($3\sigma$ depth estimated over an area of 10$\times$10 arcsec$^2$). \par 
Five galaxies out of the 24 galaxies in DECaLs need to be removed from the sample as the images suffer contamination from bright stars or overlapping sources on or near $>50\%$ of the area in the outskirts of the galaxy of interest. This criteria results in a final sample of 24 galaxies (6 in S4G, 19 in DECaLS) for the \rpeak{} and halo analysis. The rest of the 12 S4G galaxies that do not have an \rpeak{} are included in this work only for the purpose of verifying and defining the size--stellar mass relation (25+ 12 = 38 galaxies, hereafter called the CGA sample). A summary of the samples used in this work and potential biases are presented in Table \ref{tab:sample} in Appendix \ref{app:table}. \par 

\begin{figure}[ht!]
    \centering
    \includegraphics[width=\linewidth]{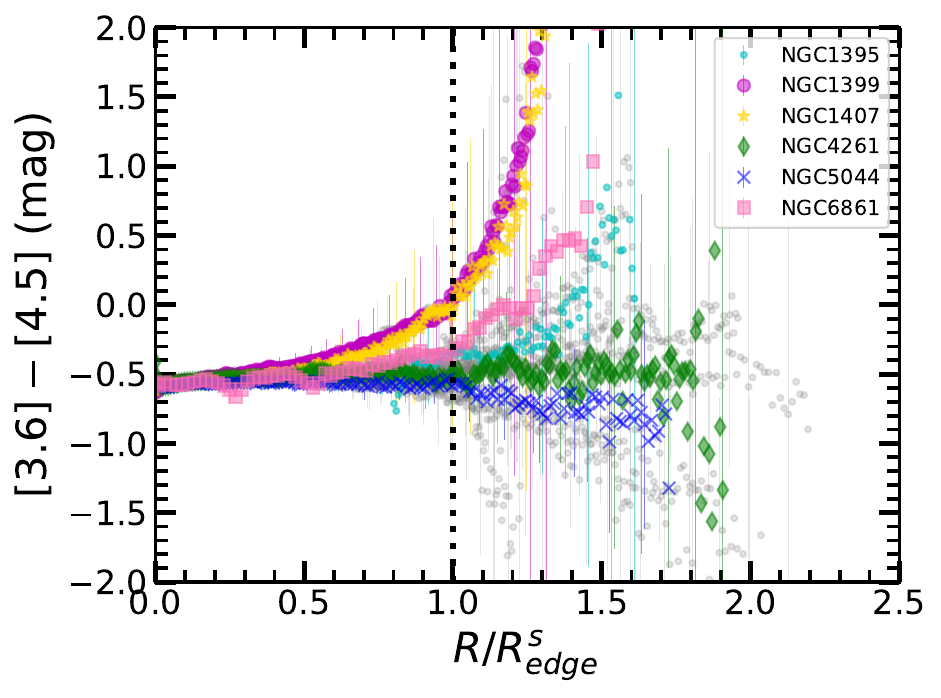}
    \caption{Mid-infrared color profiles of S4G+CGA galaxies, scaled using their identified edge radii. S4G galaxies labeled in the legend have an observable stellar edge in the color profile as well as a temperature peak  radius in their X-ray halo. Those plotted in gray do not harbor the latter. The vertical dotted line marks the feature in the color profile identified as the stellar edge.}
    \label{fig:color_profiles}
\end{figure}

While \citet{chamba+2022aap667_87, chamba+2024aap689_28} have already demonstrated the use of the g- and r-band color profiles to identify edges, following \citet{peletier+2012mnras419_2031} the technique using the mid-infrared color profiles for ETGs is demonstrated in Fig. \ref{fig:color_profiles}. The mid-infrared color is defined as [3.6] - [4.5] using the Spitzer IRAC 3.6$\mu$m and 4.5$\mu$m bands from S4G. Compared to optical wavelengths, the mid-infrared color is a much better approximation for the stellar mass distribution in galaxies with old stellar populations like ETGs, giving an uncertainty in the mass-to-light ratio of only $\sim$0.07\,dex \citep[see][]{meidt+2014apj788_144}. The lower uncertainty is due to the fact that the mid-infrared color is essentially free from dust extinction. However, optical colors can still be used in the LSB regime ($\mu_g < 26-27$~\magarc) in galaxies as dust is often found in the concentrated inner regions of galaxies while the edges are in the outskirts \citep[see also e.g.][]{dale+2016aj151_4}. \par
Fig. \ref{fig:color_profiles} shows the mid-infrared color profiles which are scaled using the edge radii (vertical dotted line). The main sources of uncertainty in both the $R^S_{\rm edge}$ and \rpeak{} measurements used in this work from the color and temperature profiles respectively are discussed in Appendix \ref{app:upper_lims}. Using these measurements, in Sect. \ref{sect:results}, we demonstrate that the edge scaling relations derived from the optical and mid-infrared technique are well-within errors and can be used to study the distribution of \rpeak{} in the CGA sample.

Notice that the shape of the color profiles shown in Fig. \ref{fig:color_profiles} are not the same for all galaxies. On the one hand, the CGA galaxies with redder colors beyond their stellar size are dominant galaxies in groups or clusters. In such cases, redder colors are expected in the outskirts given that galaxies in higher density environments have been quenched far earlier compared to nearly isolated galaxies \citep[see][]{chamba+2024aap689_28}. On the other hand, bluer colors beyond the stellar size are found in field ETGs and represent more recent accreted material \citep[][]{chamba+2022aap667_87}. Profiles with flatter slopes beyond their edge are likely those more impacted by mergers \citep[e.g.][]{borlaff+2017aap604_119}. Similar trends in the color profile of ETGs have also been published previously by other teams \citep[e.g.][]{capaccioli+2015aap581_10, spavone+2017aap603_38}. \par

\subsection{X-ray Halo Asymmetry with SAUNAS/Chandra}
\label{sect:saunas}

\begin{figure*}
    \centering
    \includegraphics[width=\linewidth]{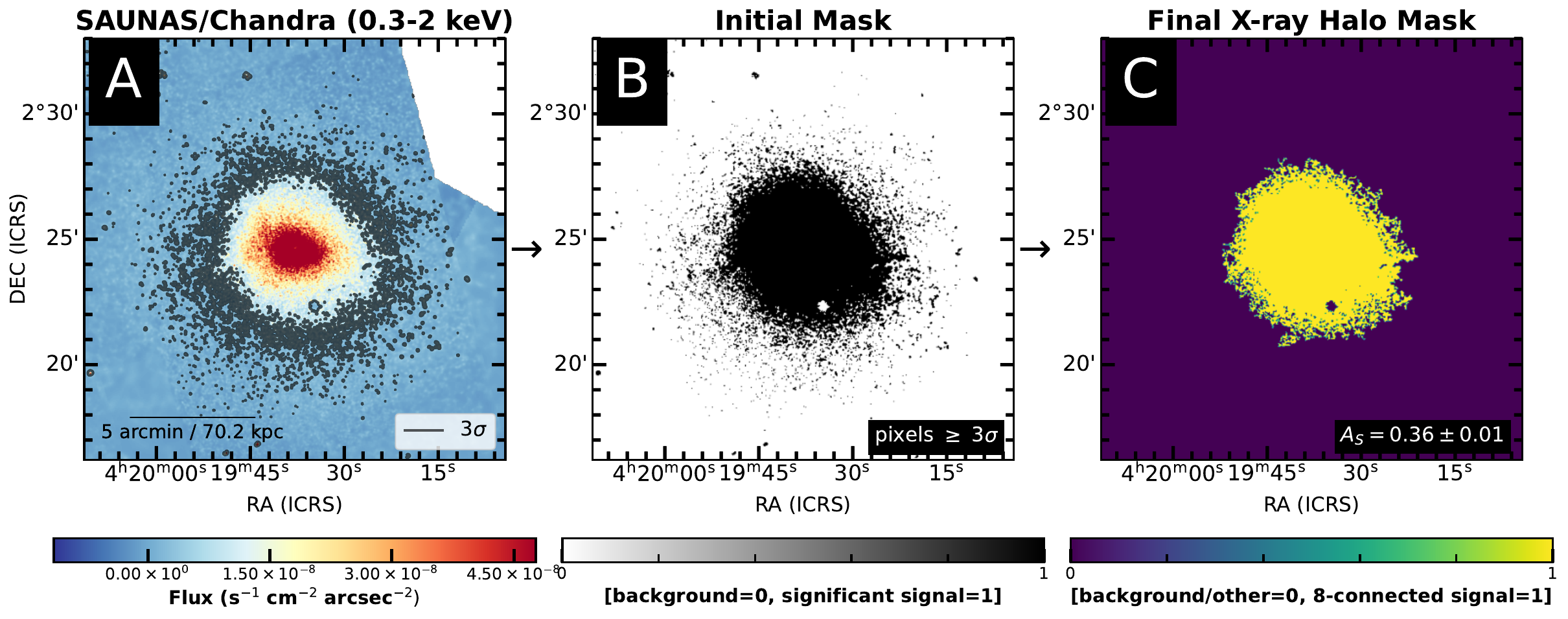}
    \caption{Selection of X-ray halo signal from the data products of SAUNAS/\emph{Chandra} and the measurement of halo shape asymmetry ($A_S$). Galaxy NGC~1550, which has one of the most massive halos in our sample, is used as an illustrative example. The method consists of three steps shown in Panels A-C. \emph{\textbf{Panel A}}: SAUNAS X-ray flux map over the energy range 0.3-2\,keV. The $3\sigma$ contour is over-plotted to highlight regions of the map that enclose significant signal. The scale bar shown in the image (3\,arcmin) is approximately the scale of the X-ray peak temperature radius (\rpeak{}) for this galaxy. \emph{\textbf{Panel B}}: Only pixels with flux greater than the 3$\sigma$ significance level from SAUNAS (A) are selected for further processing. These pixels are used to create an initial mask of the image where background is labeled as 0 and all other significant signal as 1. \emph{\textbf{Panel C}}: The 8-connectivity operator is applied to this initial mask (B) in order to isolate the X-ray halo of the galaxy as a single connected component. This component is labeled as 1 (yellow) while all other regions of the SAUNAS map comprising of background or other sources are set to 0 (purple). For example, the apparent southern hole  in purple within the yellow region is caused by the point-source removal (masking) procedure in the SAUNAS pipeline. Regions labeled as 1 (yellow regions) are used exclusively to estimate the X-ray halo shape asymmetry $A_S$. The value of $A_S$ and its uncertainty due to the binary mask is provided in the lower right of panel C. }
    \label{fig:masking_connectivity}

\end{figure*}


Finally, to quantify the asymmetry or disturbance in the X-ray halo morphology of galaxies, diffuse X-ray surface brightness maps are derived using the \textbf{S}elective \textbf{A}mplification of \textbf{U}ltra \textbf{N}oisy \textbf{A}stronomical \textbf{S}ignal \citep[SAUNAS;][]{borlaff+2024apj967_169} pipeline. In short, SAUNAS  aims to detect very faint X-ray emission in the outskirts of galaxies in order to characterize the physical properties of hot halos. \par 
The SAUNAS pipeline (1) retrieves all available observations from the \emph{Chandra} archive; (2) pre-calibrates them using well-known tools such as CIAO and MARX; (3) generates the point spread function (PSF) to account for scattered light contamination from background and point sources; and finally (4) uses LIRA \citep{donath+2022inproceedings_98} and VorBin \citep{cappellari+2003mnras342_345} to create a deconvolved, tesselated soft-band flux map which is accompanied by maps of the associated standard deviation of the flux and signal-to-noise ratio (SNR). The FoV as well as the bin size which sets the spatial scale of the resulting maps are chosen by the user. \par

The X-ray background in the images are estimated using a constant background model after all the sources in the FoV are completely masked. This choice strictly limits the use of SAUNAS only to galaxies in the \emph{Chandra} archive where the detector is not contaminated from larger background gradients \citep[see also][]{anderson+2011apj737_22}. Such gradients can arise from neighboring large X-ray emitting sources such as clusters. Only two galaxies in our sample are potentially impacted by the cluster halo \citep[NGC~4374;][]{finoguenov+2008apj686_911} and \citep[NGC~1407;][]{su+2014apj786_152}. See \citet{borlaff+2024apj967_169} for a detailed description of the SAUNAS algorithm. \par 

The SAUNAS pipeline is run for the \rpeak{} sample in the X-ray energy band range 0.3-2\,keV on a $17'\times17'$ FoV centered on each galaxy. The archival datasets used as input for SAUNAS for the full sample are listed in Appendix Table \ref{tab:chandra_archive_datasets}-\ref{tab:chandra_archive_datasets_3}. The 0.3-2\,keV energy band is selected given its use in well-known past efforts of X-ray halo characterization \citep[see e.g.][]{diehl+2007apj668_150, diehl+2008apj680_897, goulding+2016apj826_167}. Thanks to \emph{Chandra}'s high spatial resolution, maps using bin sizes of 1.97'' and 7.87'' are created in order to visualize both the small and large-scale, faint and diffuse structures of the X-ray halos. The 1.97'' (or 2'') maps are used for the main purpose of measuring the X-ray halo shape asymmetry. In Appendix \ref{app:upper_lims}, we verify that the location of \rpeak{} used are within the halo detections from SAUNAS. \par

Following \citet{pawlik+2016mnras456_3032}, shape asymmetry is defined as

\begin{equation}
\label{eq:1}
    A_S = \frac{\Sigma_{i, j} |S_{i, j} - S_{i, j}^{180}|}{\Sigma_{i, j} |S_{i, j}|}
\end{equation}

where $(i, j)$ are the pixel indices, $S$ is a binary detection mask (i.e. a segmentation map) and  $S^{180}$ is the same mask rotated by 180\,degrees about the center of the galaxy. A binary detection mask or image segmentation separates the pixels of the image in to background (labeled using zeros) and signal (labeled using ones) \citep[][]{haralick+1985inproceedings_2}. For this reason, compared to standard rotational asymmetry which depends on the flux of each pixel \citep[e.g.][]{conselice+2000apj529_886}, a background term does not appear in Eq. \ref{eq:1} as it is set to zero in the binary detection mask by construction (see \citet[][Sect. 3.3]{pawlik+2016mnras456_3032} and \citet[\texttt{statmorph};][Sect. 4.7.2]{rodriguezgomez+2019mnras483_4140}). The shape asymmetry parameter is used given that it is independent of flux and treats all pixels in the mask equally. Consequently, the measure is sensitive towards the asymmetry in the morphology of LSB detections included in the mask, rather than asymmetry due to flux distribution  \citep[see the \texttt{statmorph} package;][]{rodriguezgomez+2019mnras483_4140}. With the above definition, the value of $A_S$ mathematically ranges between 0 (perfectly symmetric) to 2 (perfectly asymmetric). However, as noted in \citet{conselice+2000apj529_886}, at least using optical imaging, values greater than one are rare in galaxies. \par 

To apply shape asymmetry in the X-ray domain, a binary detection mask for each galaxy X-ray halo is created following three steps. These steps are illustrated in Fig. \ref{fig:masking_connectivity} considering NGC~1550, a galaxy with a large X-ray halo. First, all detections in the SAUNAS 2'' map with SNR $\geq 3$ (i.e. above $3\sigma$) are selected. X-ray detections in the SAUNAS surface brightness map above this threshold are considered significant signal. Over 2'' scales, the average X-ray surface brightness in the energy range 0.3-2\,keV at this 3$\sigma$ contour for galaxies in our sample is between 1.4$\pm0.4 \times10^{-9}$ and 8.5$\pm3.5\times10^{-9}$ s$^{-1}$ cm$^{-2}$ arcsec$^{-2}$. These values are faint enough to include LSB features in the X-ray domain (see Introduction). An initial mask is created where all pixels above $3\sigma$ are set to one, while the rest of the pixels are considered as `background' and are set to zero. \par 
Second, the 8-connectivity rule in image processing\footnote{See ``An Introduction to Digital Image Processing'' (Roedink, 1998): \href{https://www.cs.rug.nl/~roe/publications/IntroDig_Iman1998.pdf}{link}. \citet{pawlik+2016mnras456_3032} used this same operator to create the binary detection mask needed to measure shape asymmetry.} is applied to these 3$\sigma$ significant detections in order group the pixels belonging to the galaxy. 8-connectivity specifically groups each pixel with their immediate neighboring pixels in all directions , i.e. a $3\times3$ matrix whose central pixel is connected in all four of its sides and edges. This operation assembles the connected components within the initial mask and defines the sets of pixels in SAUNAS which are linked and form objects. \par 
Finally, all other pixels not belonging to the galaxy connected component are masked by setting them to zero. The galaxy X-ray halo connected component is selected by using the central coordinates of the galaxy which are listed in Tables \ref{tab:chandra_archive_datasets}-\ref{tab:chandra_archive_datasets_3}. This final step results in the final X-ray halo mask which defines the region of the image used to measure the X-ray halo shape asymmetry.\par

Given that $A_S$ depends on the rotated mask $S^{180}$, the coordinates of the center of the galaxy must be chosen carefully in order to avoid artificially high asymmetric values (i.e. false positives). To compare the location of the X-ray peak temperature with respect to the (optically derived) stellar size--stellar mass plane, the flux-weighted optical center of the galaxy must be used as the common reference point. Following  \citet{rodriguezgomez+2019mnras483_4140} (\texttt{statmorph}), the optical center of the galaxy is chosen such that the standard rotational asymmetry applied to the flux image is minimized. This minimization criteria ensures that the rotation is performed about the core of the galaxy. The central coordinates derived from this procedure using the \texttt{statmorph} package for each galaxy using DECaLS or S4G imaging (whichever available for the sample, see Sect. \ref{sect:size_plane}) is listed in Table \ref{tab:chandra_archive_datasets}-\ref{tab:chandra_archive_datasets_3}.

The main source of uncertainty in the value of $A_S$ for each galaxy arises from the uncertainty in the binary detection mask. The uncertainty in the mask depends on three factors: (1) the chosen threshold used to derive the binary mask, (2) the resolution of the data and (3) the noise in the SAUNAS surface brightness map.  

The dependence of the mask on the chosen resolution and thresholding have been discussed extensively elsewhere by other authors \citep[see][]{pawlik+2016mnras456_3032, sazonova+2024afz7_77}. Following those works, Appendix \ref{app:a_s_uncert} shows that the results discussed here are valid at a spatial scale of 0.57$\pm0.3$\,kpc (or 2'' across the full sample) and the choice of threshold used is limited by the depth and noise of the \emph{Chandra} data. A threshold lower than 3$\sigma$ artificially increases the value of $A_S$ in all galaxies. This increase occurs because lowering the threshold includes areas of the image with lower detection significance in the final the binary detection mask (see Appendix \ref{app:a_s_uncert}). As the choice of the 3$\sigma$ threshold to derive the binary mask is based on the SAUNAS signal-to-noise map (Sect. \ref{sect:orign_halo_asym}), the uncertainty in $A_S$ can be estimated using a monte carlo approach. The procedure used to estimate the uncertainty in the $A_S$ value due to the noise and 3$\sigma$ threshold in the SAUNAS data products for each galaxy is also explained in Appendix \ref{app:a_s_uncert}. On average, the total uncertainty in $A_S$ is in the order of 3\%.

\begin{figure*}
    \centering
    \includegraphics[width=0.46\linewidth]{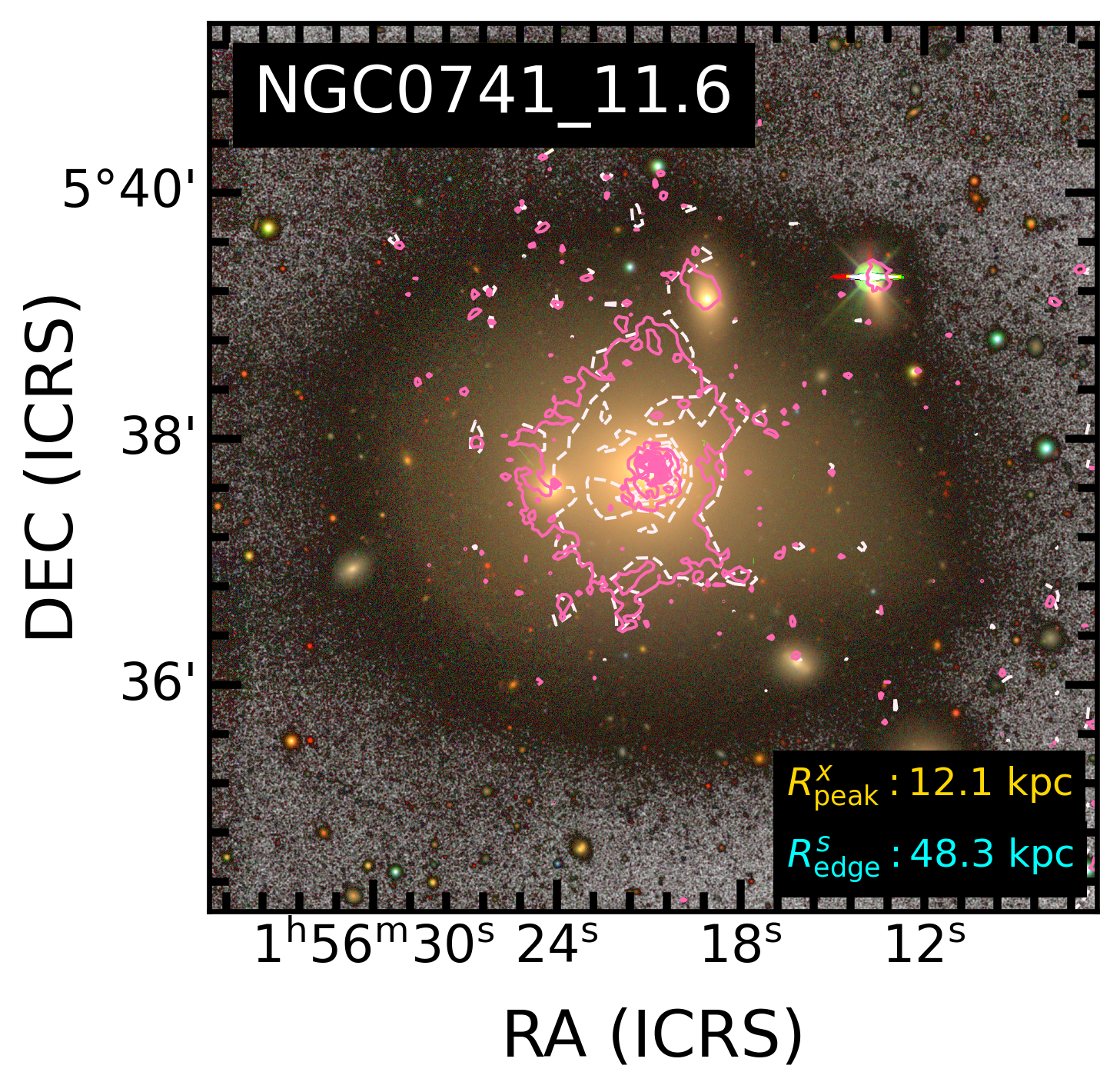} \hspace{0.4cm}
    \includegraphics[width=0.5\linewidth]{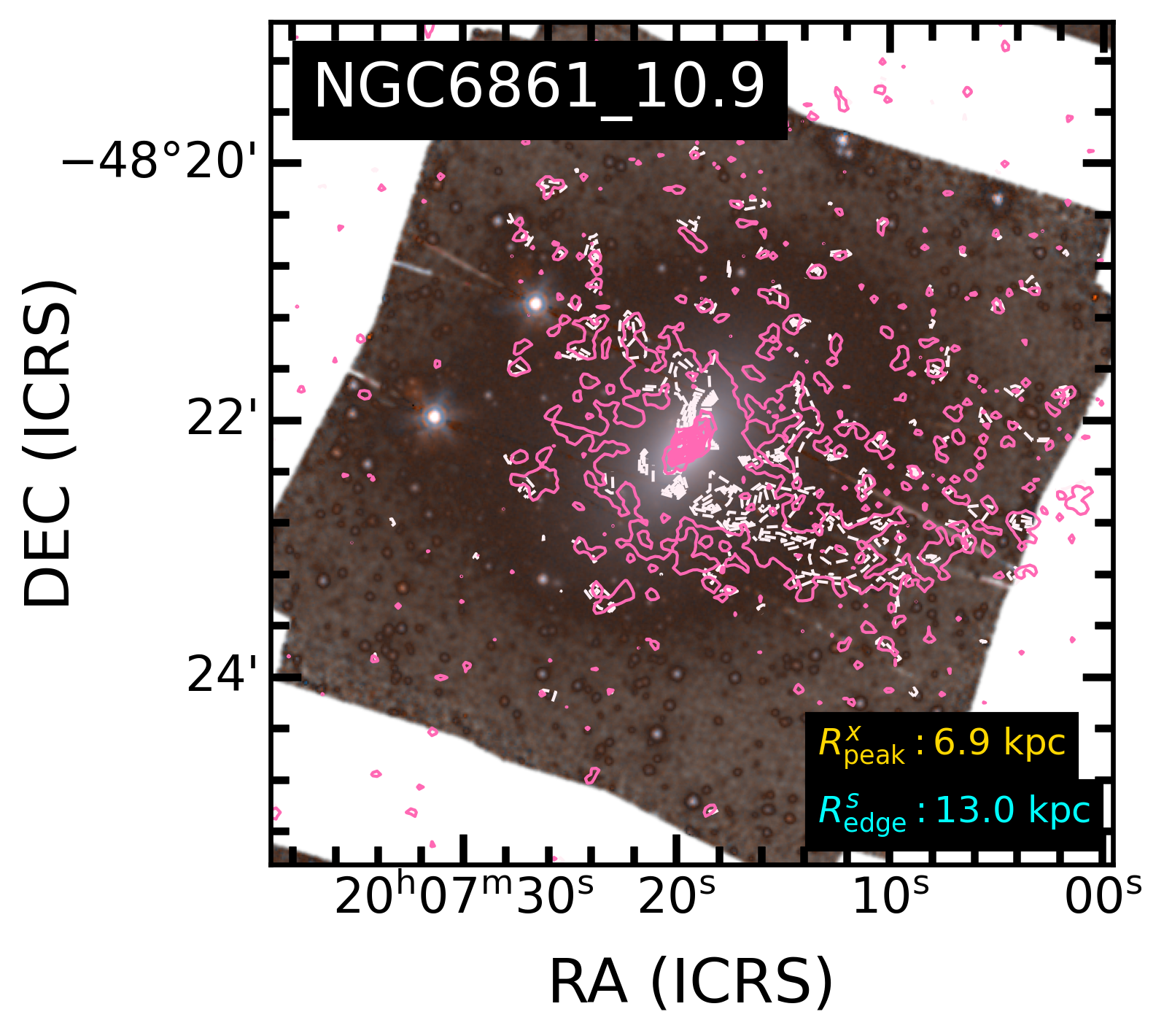} 
    \\

    \includegraphics[width=0.53\linewidth]
    {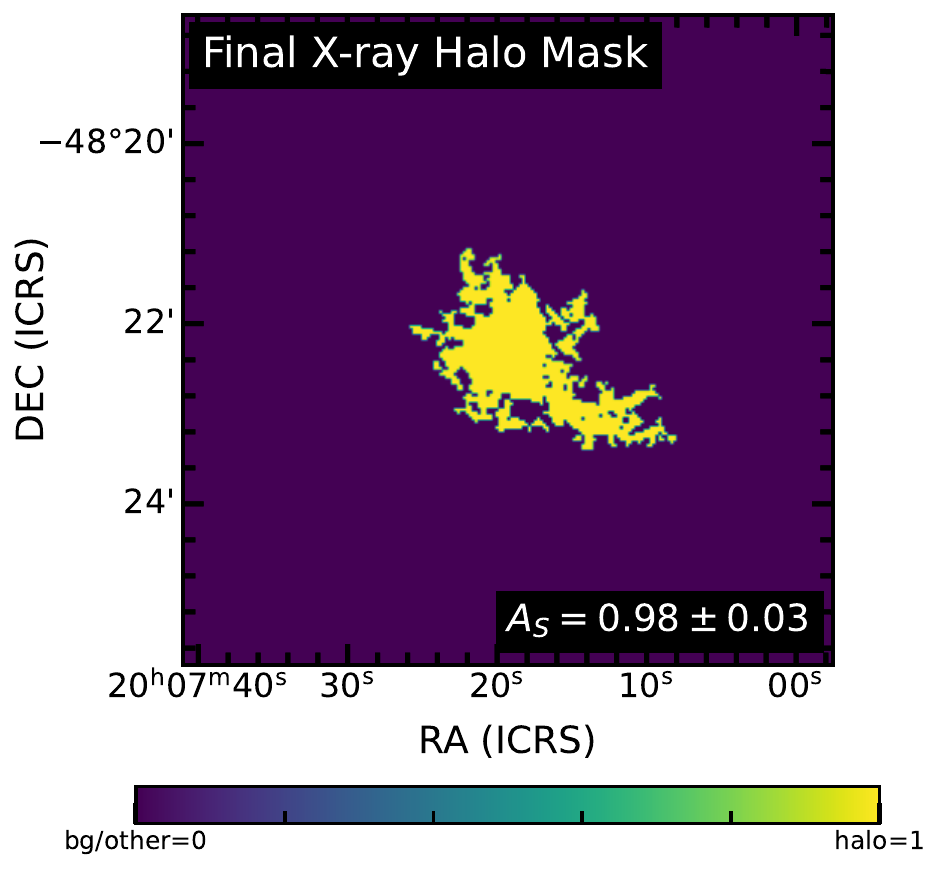}
    \hspace{-18.7cm}
    \includegraphics[width=0.5\linewidth]
    {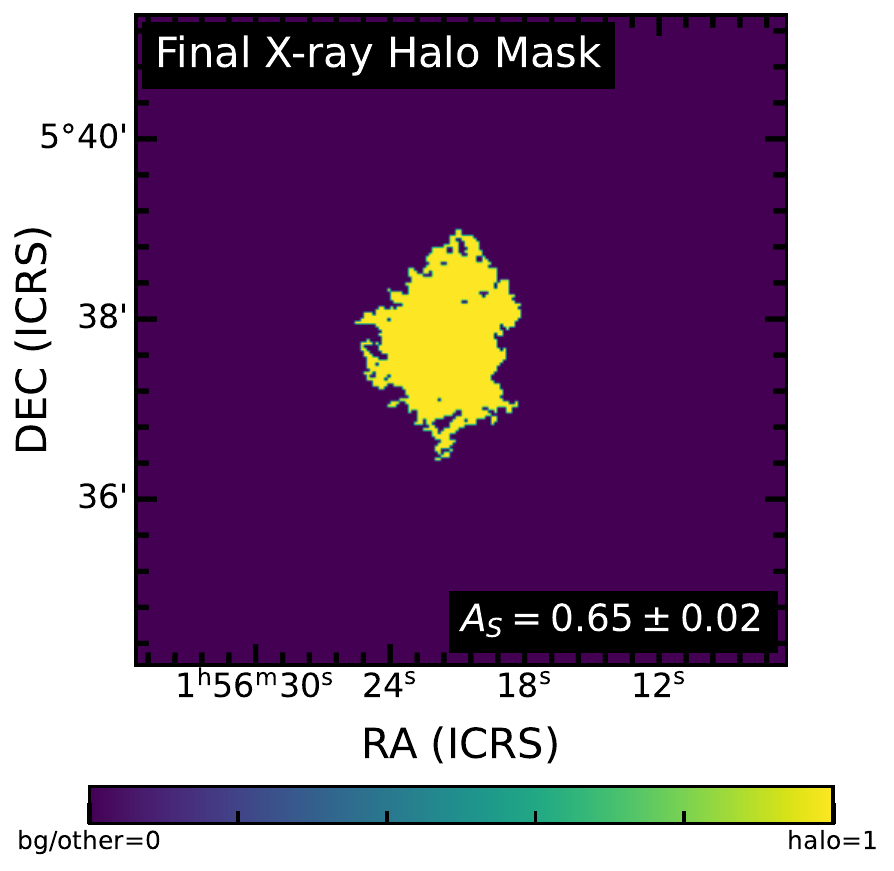}

    \caption{X-ray Halo asymmetry using SAUNAS/\emph{Chandra} for galaxies NGC~0741 (left) and NGC~6861 (right) as illustrative examples of more asymmetric halos compared to NGC~1550 from the previous Fig. \ref{fig:masking_connectivity}. \textit{\textbf{Upper}}:  The location of $R^x_{\rm peak}$ (yellow solid ellipse) and $R^s_{\rm edge}$ (cyan dotted) are over-plotted on combined optical images with the background colored using a gray scale (NGC~0741 using DECaLS, and NGC~6861 using S4G). The full range of the soft-band surface brightness map from SAUNAS is also plotted as 20 contours (pink) which are logarithmically spaced between the signal-to-noise per pixel range of 3 to the maximum detected in SAUNAS for each galaxy. Each panel is labeled using a pattern based on the common name of the galaxy and rounded stellar mass used in this work, i.e. name\_mass. \textit{\textbf{Lower}}: Final X-ray halo mask used to estimate the asymmetry of the X-ray halo shape (Eq. \ref{eq:1}). As in Fig. \ref{fig:masking_connectivity}, pixels considered part of the X-ray halo are labeled as 1 (yellow) while all other pixels belonging to the background or other sources from SAUNAS are set to 0 (purple). The $A_S$ value obtained in each case is included in the lower part of the panels. Increasing $A_S$ values correspond to visually more asymmetric X-ray halos.}

    \label{fig:categories}
\end{figure*}

Two more examples of the SAUNAS surface brightness and associated binary detection maps are shown in Fig. \ref{fig:categories}. The resulting value of $A_S$ from Eq. \ref{eq:1} are also included in the figure. The maps show that increasing $A_S$ values correspond to visually more asymmetric X-ray halos. The above methods for analyzing the distribution of \rpeak{} and $A_S$ in the stellar size--stellar mass plane are applied to the rest of the sample and are presented in the next section.

\section{Results}
\label{sect:results}

\begin{figure*}[h!]
    \centering

    \includegraphics[width=0.9\linewidth]{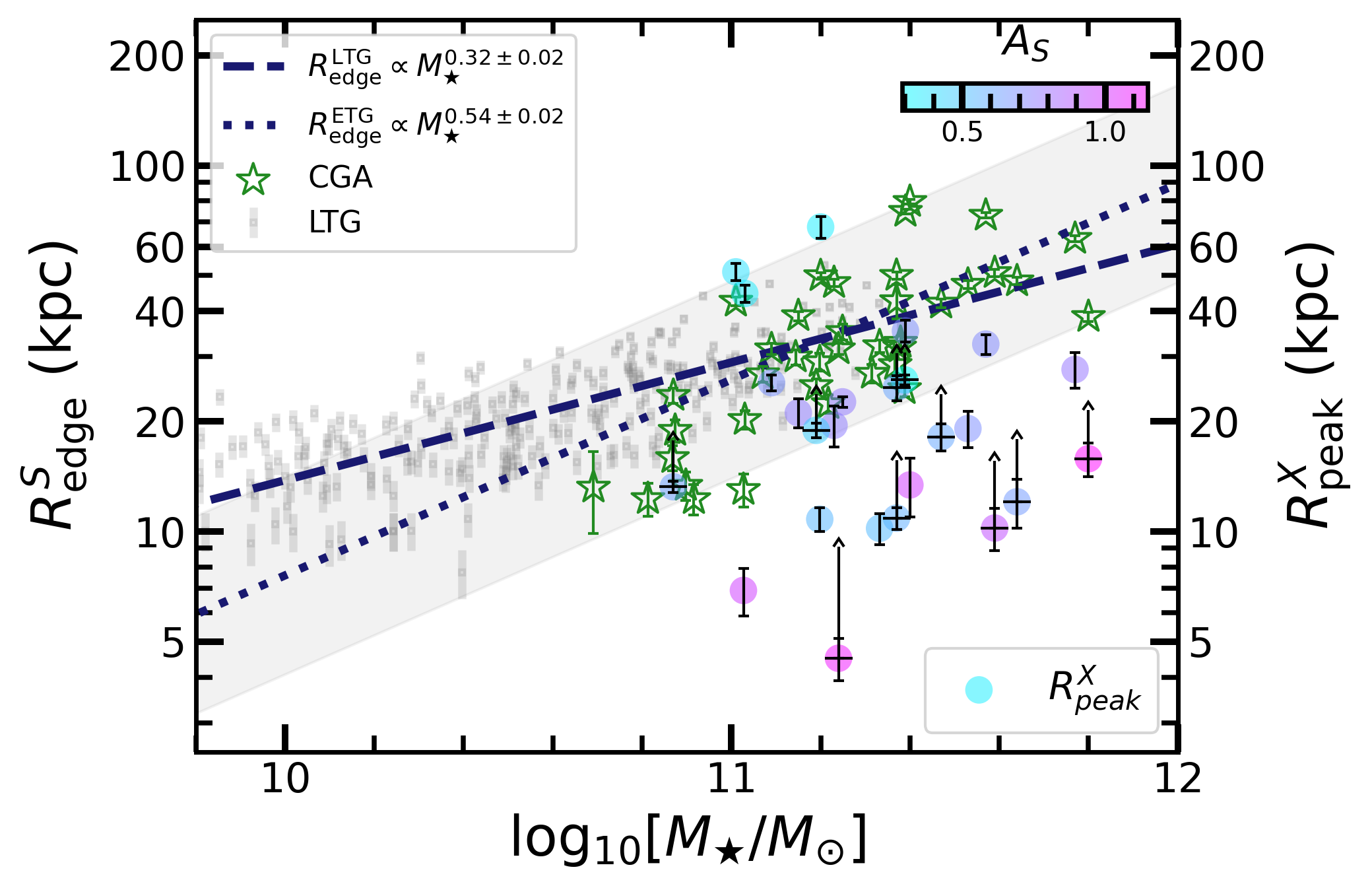}

    \caption{Distribution of $R^S_{\rm edge}$ and \rpeak{} as a function of stellar mass $M_{\star}$. The $R^S_{\rm edge}$ of the CGA sample identified in this work are plotted as green stars. The $R^S_{\rm edge}$ relations for late-type and early-type field galaxies (LTG and ETG, respectively) from \citet{chamba+2022aap667_87} are plotted using dashed and dotted lines. For legibility, only those individual measurements used to derive the best fit LTG dashed line (small gray squares) are shown. The shaded area represents the $3\sigma$ scatter around dotted ETG relation. The \rpeak{} values are over-plotted in the same $R^S_{\rm edge}-M_{\star}$ plane (points) and are colored according to the shape asymmetry of the galaxy's X-ray halo ($A_S$, see text for details). Larger $A_S$ values indicate more asymmetry. The total uncertainties in each measurement are described in Appendix \ref{app:upper_lims}. The galaxies with upward arrows on their \rpeak{} value are the ten cases identified as lower limits in this work because their original \rpeak{} values lay beyond SAUNAS's threshold SNR radius (see Sect. \ref{sect:saunas} and Appendix \ref{app:upper_lims}). All galaxies laying below the 3$\sigma$ shaded area show significant asymmetry  in their X-ray halo morphology.}
    \label{fig:edges_peak_radii}
\end{figure*}

The stellar size--stellar mass plane in shown in Fig. \ref{fig:edges_peak_radii}. The best fit lines defined by the slope and intercept for the CGA (green stars, 38 galaxies) and ETG sample from \citet{chamba+2022aap667_87} individually are well within errors. The LTG relation (dashed) highlights that the ETGs, including the CGA sample, follow a distinct size-mass relation \citep[see][]{trujillo+2020mnras493_87, chamba+2022aap667_87}. This result justifies the use of the $R^{ETG}_{\rm edge}$ relation in our analysis and is plotted as a dotted line. The radial location of the X-ray peak temperature values, \rpeak{}, are plotted as a function of $M_{\star}$ in the same plane and color coded according to their X-ray halo asymmetry $A_S$ value. \par

\begin{figure*}
    \centering
    \includegraphics[width=0.49\linewidth]{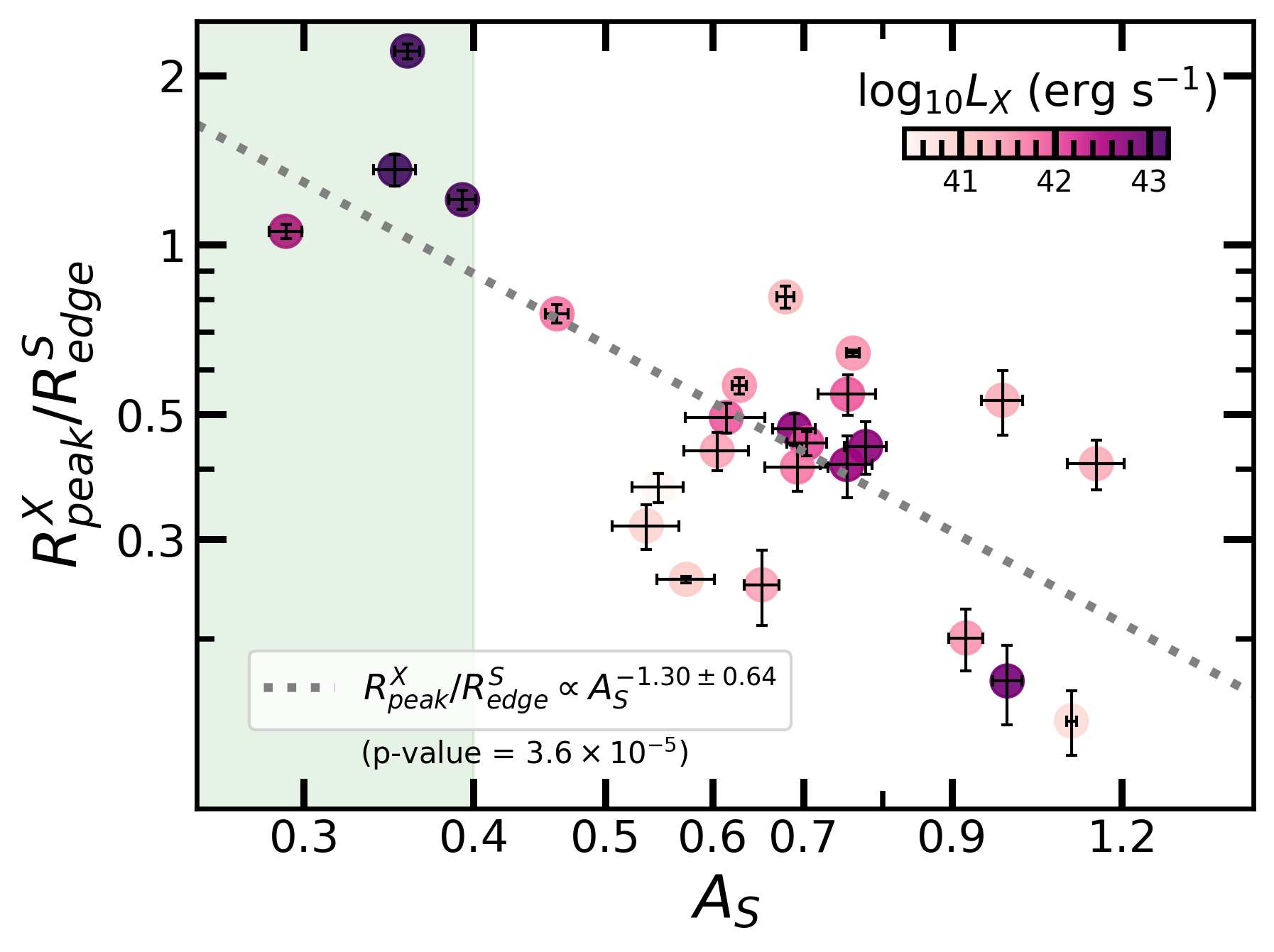}
    \includegraphics[width=0.49\linewidth]{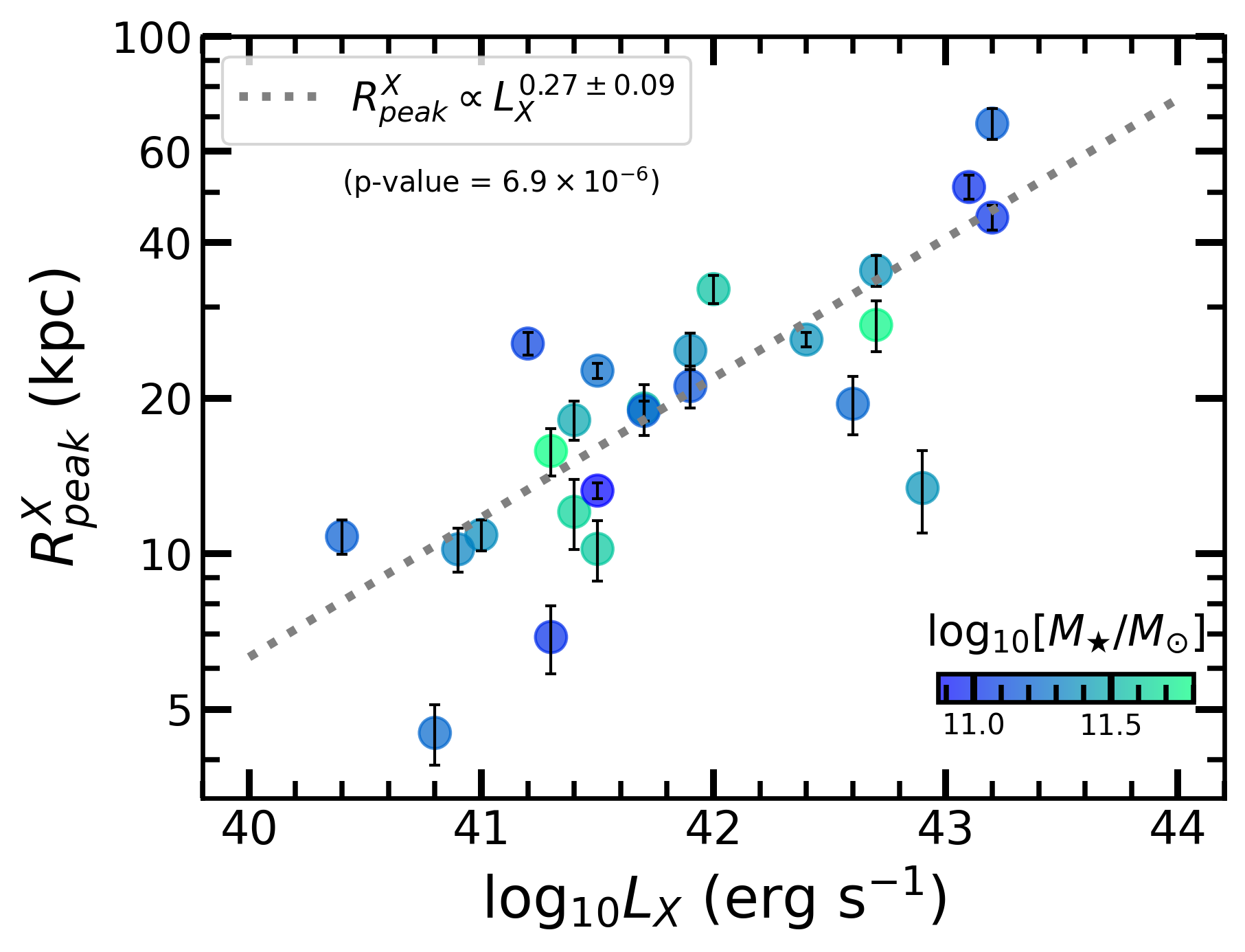}
    \caption{Correlations between the peak X-ray temperature radius \rpeak{}, X-ray halo luminosity $L_X$ and halo shape asymmetry $A_S$. \textbf{\emph{Left:}} The ratio \rpeak{}/$R^S_{\rm edge}$ plotted as a function of $A_S$. The regime where $A_S < 0.4$ is shaded light green to highlight the lack of galaxies in our sample with low asymmetries and large \rpeak{}. Each data point is color coded according to the galaxy's $L_X$. \textbf{\emph{Right:}} \rpeak{} as a function of $L_X$. Each data point is color coded according to the galaxy's stellar mass $M_{\star}$. Both correlations are statistically significant with p-values in the order of 10$^{-5}$ and 10$^{-6}$. The figure shows that galaxies with more truncated (or small) \rpeak{} with respect to their stellar boundaries $R^S_{edge}$ have more asymmetric X-ray halos. Despite this dependence on asymmetry, \rpeak{} is tightly correlated with $L_X$. The physical significance of these correlations are discussed in Sect. \ref{sect:correlation}.}
    \label{fig:rpeak_corr}
\end{figure*}

In Fig. \ref{fig:rpeak_corr}, the left panel displays the correlation between the ratio \rpeak{}$/R^S_{\rm edge}$ and X-ray halo shape asymmetry $A_S$ with each data point color coded according to the X-ray luminosity of the galaxy halo $L_X$. The regime where $A_S < 0.4$ is shaded light green to highlight the lack of galaxies in our sample with low asymmetries and large \rpeak{}. The right panel shows the relationship between \rpeak{} and $L_X$ with the data points color coded using the galaxy's stellar mass $M_{\star}$. In each panel, the best-fit linear relation is plotted as a dotted gray line. 

The main findings from these relations are: First, in Fig. \ref{fig:edges_peak_radii}, at a fixed $M_{\star}$ galaxies with small \rpeak{} have higher $A_S$ value. This result is reinforced by the strong negative correlation between the ratio \rpeak{}/$R^S_{\rm edge}$ and $A_S$ shown in Fig. \ref{fig:rpeak_corr} (Pearson correlation coefficient $r=-0.73$; p-value=$3.5\times10^{-5}$). In other words, the X-ray peak temperature radius \rpeak{} is more truncated with respect to the galaxy's  stellar boundary when the X-ray halo shape is more asymmetric, following \rpeak{}/$R^S_{\rm edge} \propto A_S^{-1.30 \pm 0.64}$. 

Second, while the correlation between \rpeak{} and $M_{\star}$ is weak with a large scatter (Fig. \ref{fig:edges_peak_radii}), the correlation between \rpeak{} and X-ray luminosity $L_X$ is strong ($r=0.77$; p-value=$6.9\times10^{-6}$; Fig. \ref{fig:rpeak_corr}), following \rpeak{}$ \propto L_X^{0.27\pm 0.09}$. The physical significance of this correlation is discussed in Section \ref{sect:discussion}.

Third, three galaxies are found with large \rpeak{}$ > R^{ETG}_{\rm edge}$ in Fig. \ref{fig:edges_peak_radii}. These galaxies are IC~1262, 1550 and 4325 and have $A_S < 0.4$, i.e their X-ray halos do not appear to have any significant asymmetry in the outermost regions (see Fig. \ref{fig:categories} and Appendix Fig. \ref{fig:no_asym}). With the exception of NGC~5044, all the other galaxies with \rpeak{}$ < R^{ETG}_{\rm edge}$ have larger $A_S > 0.4$. In other words, the majority of our sample with small \rpeak values have disturbed hot halos. The case of NGC~5044 is related to the uncertainty in \rpeak{} and discussed in more detail in Appendix \ref{app:upper_lims}. \par

\begin{figure*}
    \centering
    \includegraphics[width=0.65\linewidth]{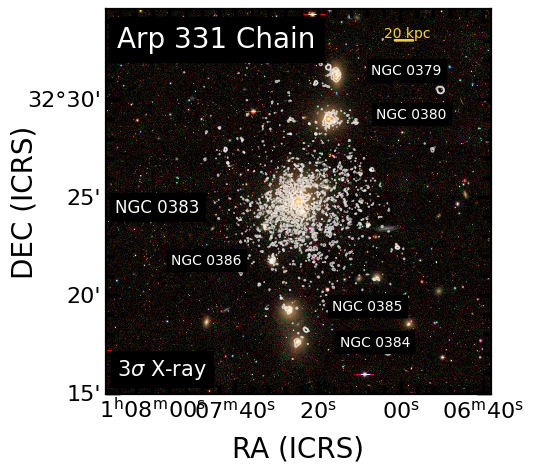}\\
    \includegraphics[width=0.57\linewidth]{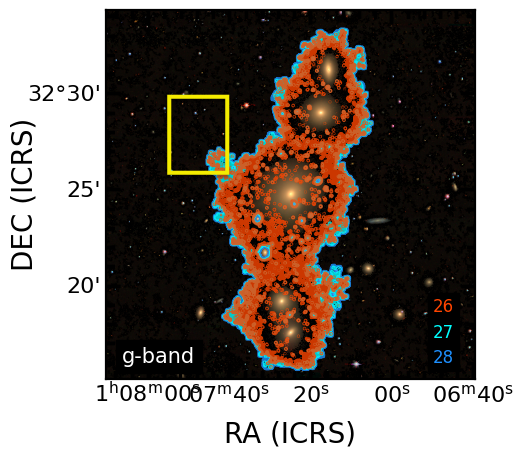}\hspace{-18.73cm}
    \includegraphics[width=0.57\linewidth]{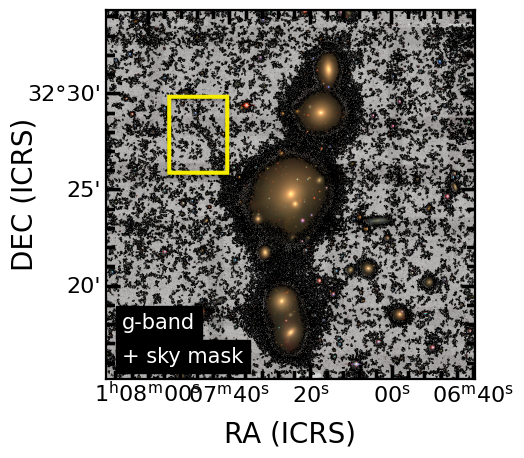} 
    
    \caption{Large stellar stream detected in NGC~0383, the central, radio galaxy in the Arp 331 Chain. The upper panel shows the main X-ray emitting galaxies in the chain with the SAUNAS/\emph{Chandra} contours over-plotted in white in the same manner as in the upper panels of Fig. \ref{fig:categories}. A zoomed view of the X-ray halo of NGC0383 is shown later in the first upper panel of Fig. \ref{fig:small_rpeak}. The lower panels visualize the intra-group light of the chain in two ways. The lower left panel shows all the detected sources in the background of NGC~0383 as black pixels. The yellow box (length $\sim 35$\,kpc) highlights a fragmented, faint feature that potentially extends the size of the stream found in the outskirts of NGC~0383 to $\sim$ 45\,kpc (see text for details). The lower right panel displays the segmentation maps of the DECaLs g-band image using Sourcerer \citep{teeninga+2016MMTA1_100, haigh+2021aap645_107}, a tool specialized for the detection of low surface brightness emission and nested sources. As a guideline, the labeled colored lines correspond to the average $g$-band surface brightness of the detected pixels.}
    \label{fig:stream_NGC0383}
\end{figure*}

\begin{figure*}[h!]
    \centering
   
    \includegraphics[width=0.54\linewidth]
    {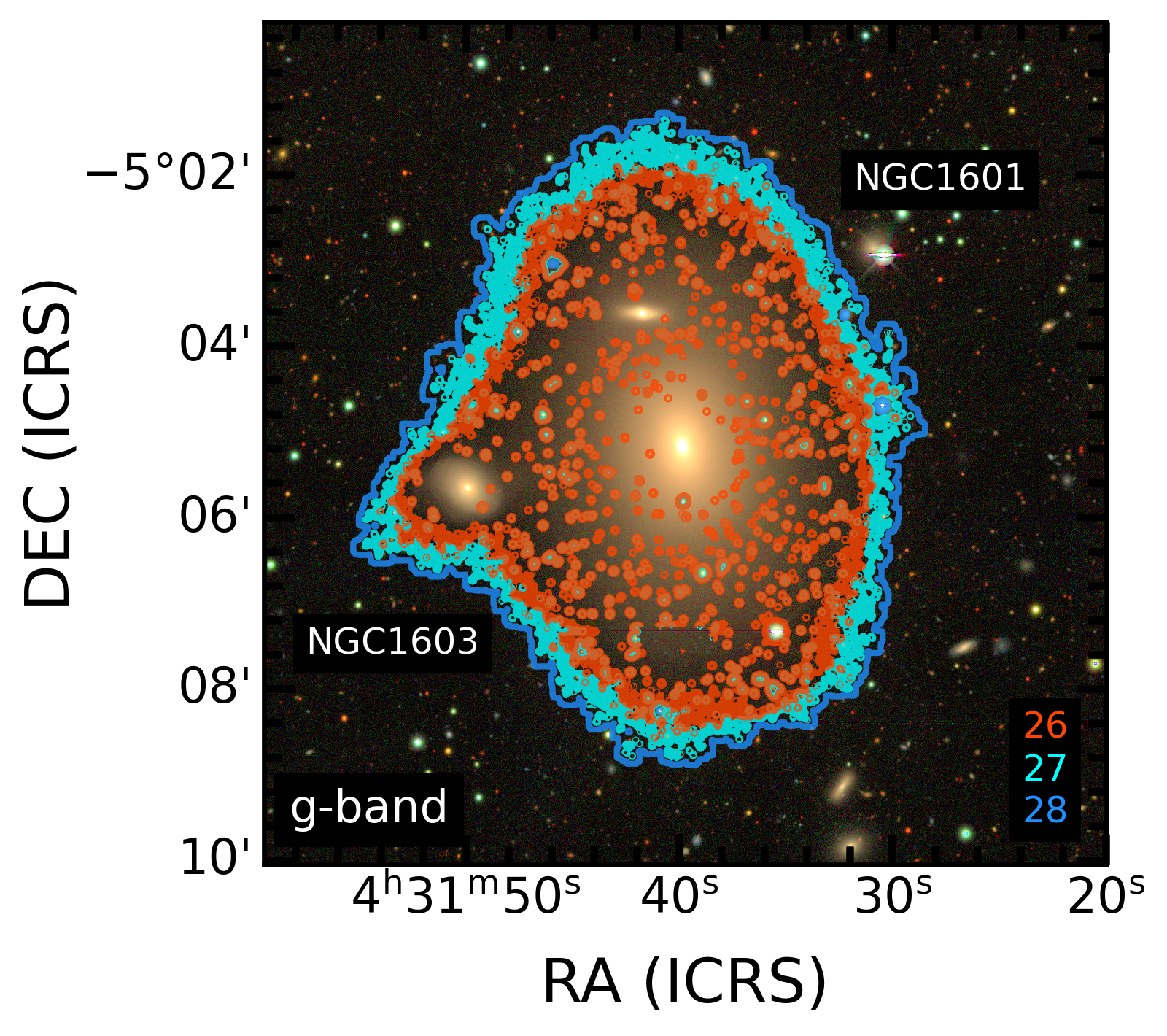}
    \hspace{-18.5cm} 
    \includegraphics[width=0.51\linewidth]{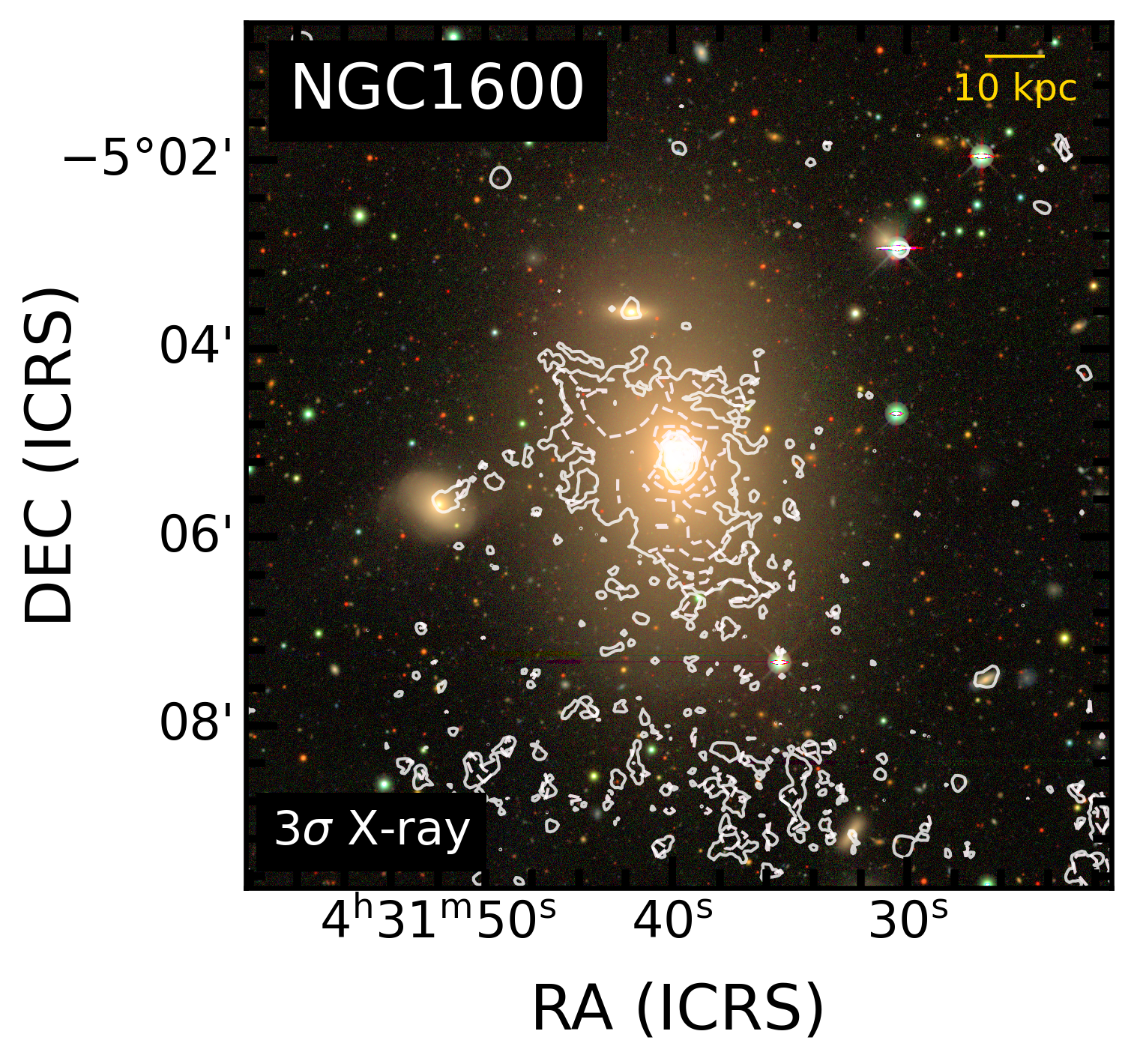}
    \\
    \includegraphics[width=0.52\linewidth]{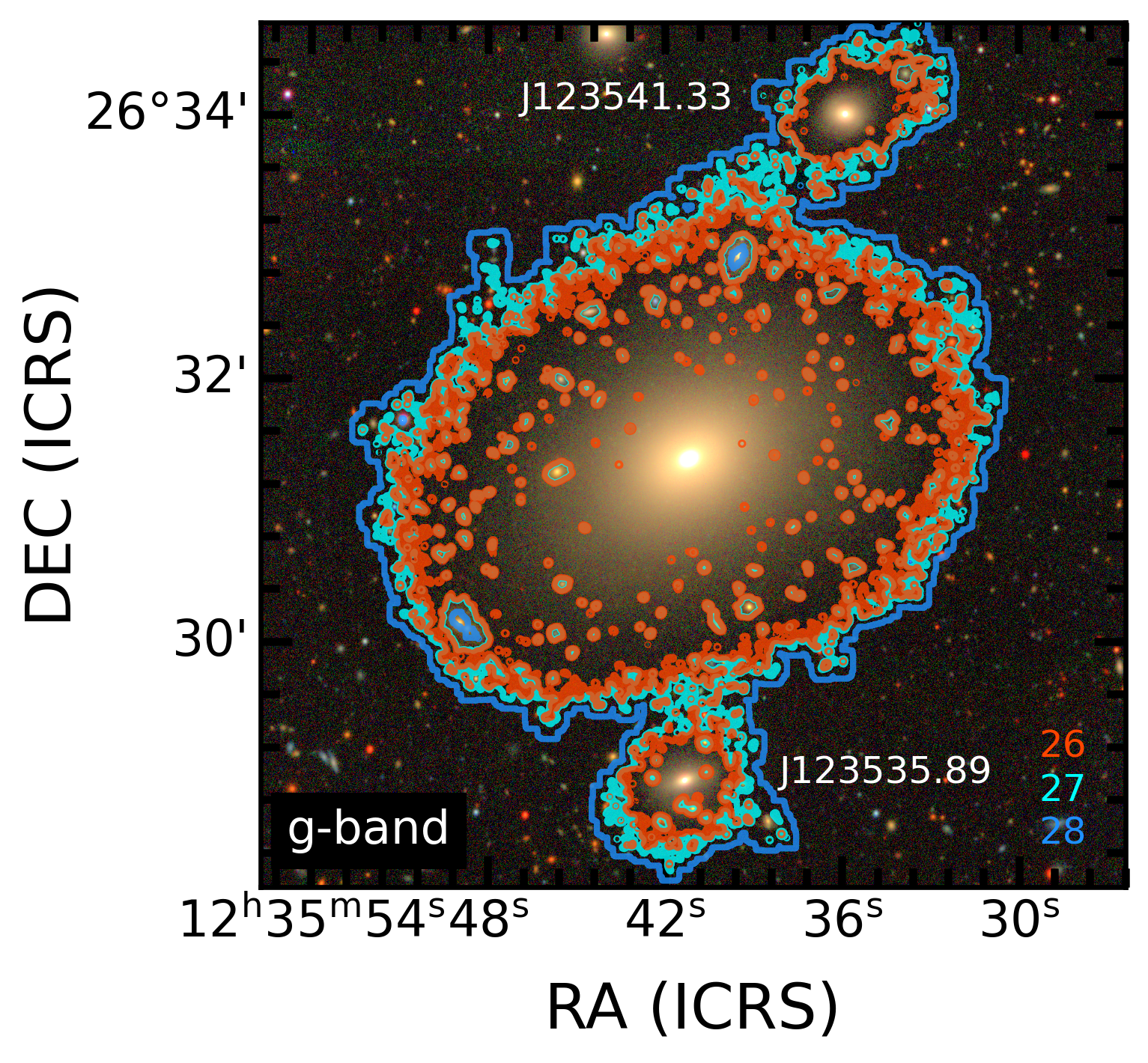}\hspace{-18cm}
    \includegraphics[width=0.51\linewidth]{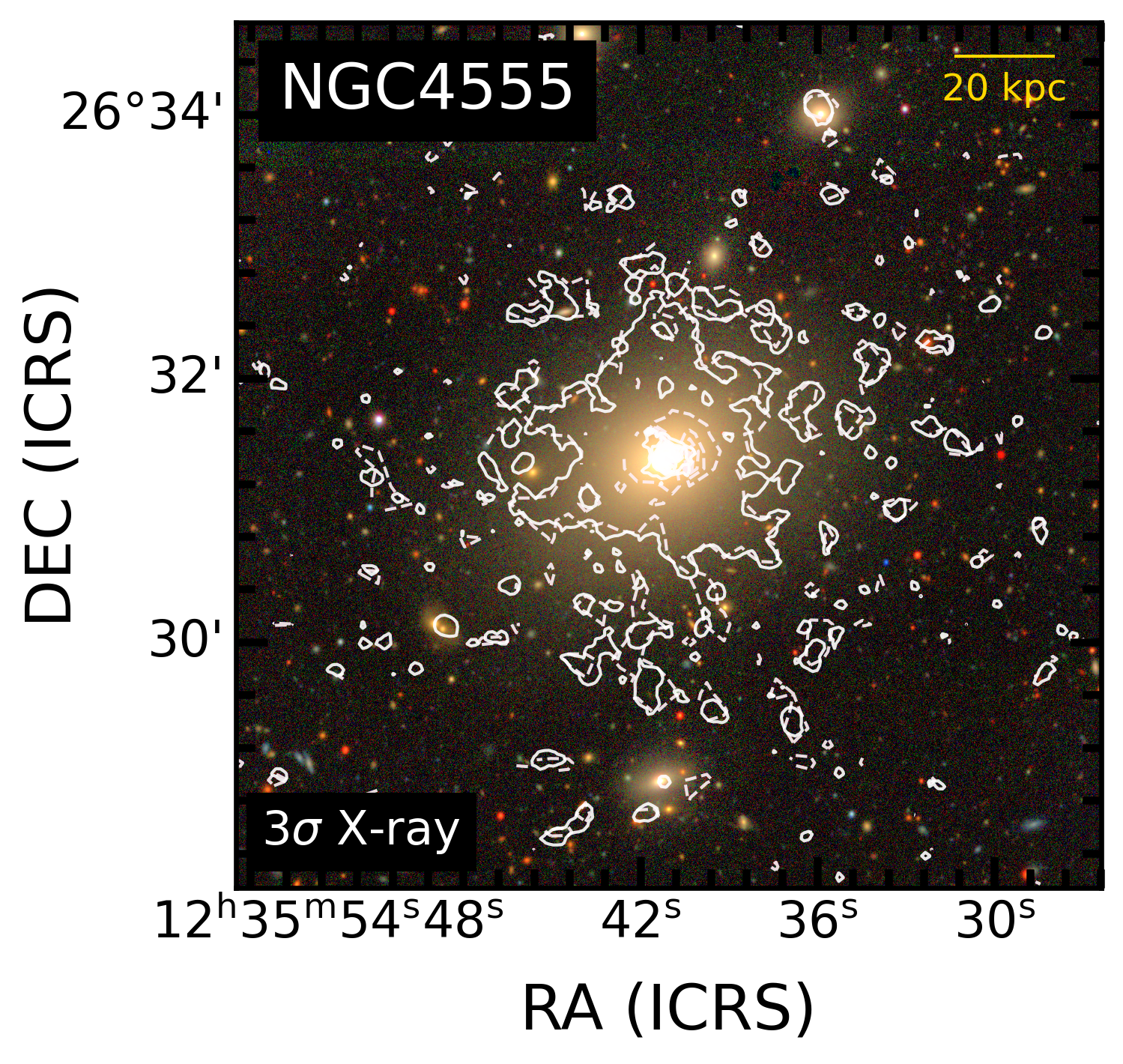}

    \caption{Discovery of disturbed X-ray halos and tidal features in ETGs NGC~1600 (upper) and NGC~4555 (lower). In both rows, the left panel displays the white contours from SAUNAS/\emph{Chandra} as in the previous figures. The size of the yellow scale bar shown in the upper right of each panel corresponds to $\sim 1$\rpeak{} of each galaxy for reference. The right panels show the segmentation maps of the DECaLs g-band image using Sourcerer \citep{teeninga+2016MMTA1_100, haigh+2021aap645_107}, a tool specialized for the detection of low surface brightness emission and nested sources. As a guideline, the labeled colored lines correspond to the average $g$-band surface brightness of the detected pixels. The nearest neighbors at a comparable redshift to the main galaxies within their FOV are also labeled. The detected features in DECaLs connecting the neighbors with the outer regions of the galaxy are signs of tidal interactions or minor mergers. How these faint optical features relate to the X-ray halo asymmetries are discussed in Sect. \ref{sect:discussion}.}
    \label{fig:ngc4555_1600}
\end{figure*}

\begin{figure*}
    \centering
    \includegraphics[width=0.32\linewidth]{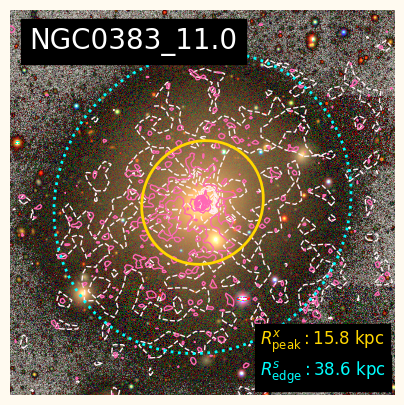} 
    \includegraphics[width=0.32\linewidth]{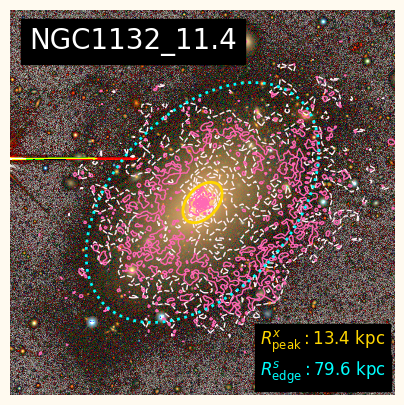}
    \includegraphics[width=0.32\linewidth]{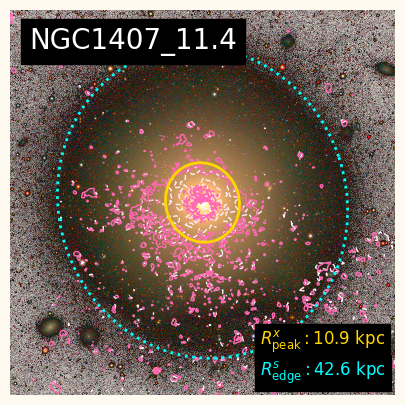}
    \includegraphics[width=0.32\linewidth]
    {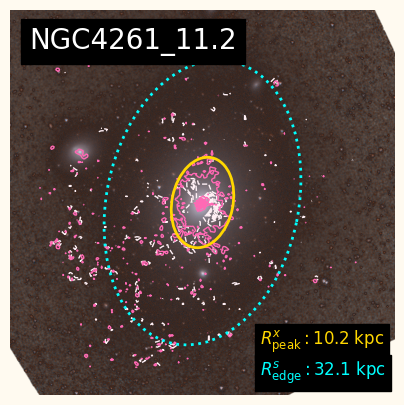}
    \includegraphics[width=0.32\linewidth]{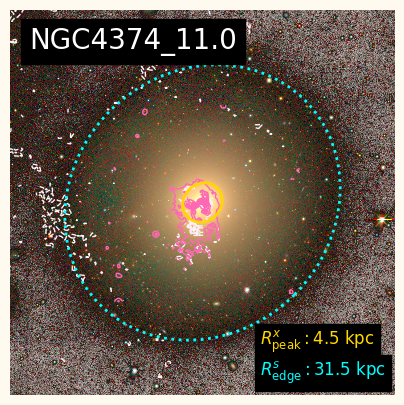}
    \includegraphics[width=0.32\linewidth]{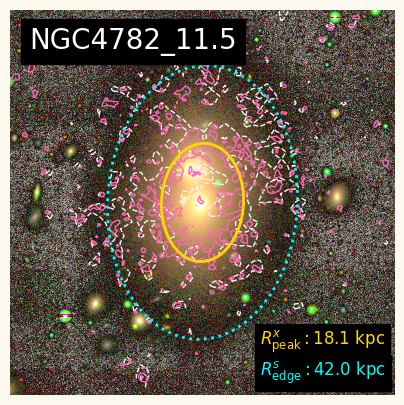}

    \caption{X-ray halos identified with very small \rpeak $<< R^{ETG}_{\rm edge}$. Each panel is annotated in the same way as the upper panels in Fig. \ref{fig:categories}. The ellipses represent the physical scale of the stellar edge (cyan, dotted) and X-ray peak temperature radius (yellow, solid) for each galaxy. All halos have a shape asymmetry $A_S > 0.4$. Galaxies NGC~0741, 6861 (Fig. \ref{fig:categories}), 1600 and 4555 shown in the previous figures also have \rpeak $<< R^{ETG}_{\rm edge}$. The central coordinates of the galaxies shown are provided in Table \ref{tab:chandra_archive_datasets}-\ref{tab:chandra_archive_datasets_3}.}
    \label{fig:small_rpeak}
\end{figure*}

Fourth, considering the galaxies with very small \rpeak{} $<< R^{ETG}_{\rm edge}$, defined as those below the shaded 3$\sigma$ $R^{ETG}_{\rm edge}$ region, we discover new X-ray and optical features in three of these galaxies, namely NGC~0383, 1600 and 4555 displayed in Fig. \ref{fig:stream_NGC0383} and \ref{fig:ngc4555_1600}. In all cases the X-ray halo morphology appears highly disturbed ($A_S \geq 0.6$), especially in the direction of the galaxy's nearest satellites. The rest of the sub-sample with \rpeak{} $<< R^{ETG}_{\rm edge}$ are shown in Fig. \ref{fig:small_rpeak}. The physical origin of the X-ray morphologies of these galaxies and their connection with \rpeak{}  are discussed in the next section.\par

\section{Discussion}
\label{sect:discussion}

This discussion is divided in to three parts. Section \ref{sect:orign_halo_asym} presents evidence to argue that very small \rpeak{} $<< R^{ETG}_{\rm edge}$ and large $A_S \geq 0.6$ in X-ray halos (Figs. \ref{fig:categories}-\ref{fig:small_rpeak}) are signatures of recent mergers or galaxy interactions. Section \ref{sect:correlation} explores how this result is compatible with the strong positive correlation between \rpeak{} and $L_X$. Section \ref{sect:new_discovery} analyzes the new halo detections presented in Fig. \ref{fig:ngc4555_1600}, considering the previous work and environments of those galaxies.

\subsection{Hot halos of galaxies with \rpeak{} $<< R^{ETG}_{\rm edge}$}
\label{sect:orign_halo_asym}

The sub-sample of galaxies where their \rpeak{} is much smaller than their stellar size $<< R^{ETG}_{\rm edge}$ are those with more X-ray halo shape asymmetry. As galaxy interactions, mergers and feedback could cause X-ray halo asymmetries, this section discusses which of these processes could have led to the highly truncated \rpeak{} in these galaxies. According to the descriptions of individual galaxies by the CGA team \citep{kim+2019apj241_36} \footnote{\protect\url{https://cxc.cfa.harvard.edu/GalaxyAtlas/Common_Fig/AppendixB.html}}, the sub-sample of galaxies where \rpeak{} $<< R^{ETG}_{\rm edge}$ (see Figs. \ref{fig:categories}--\ref{fig:small_rpeak}) consists of six galaxies which are the dominant one in their respective group environment, one cluster galaxy (namely NGC~4374), one known fossil group \citep[NGC~1132 which is disturbed both in X-rays and optical wavelengths; see Fig. \ref{fig:small_rpeak};][and references therein]{alamomartinez+2012aap546_15, kim+2018apj853_129} and two galaxies which are nearly isolated (NGC~1600 and 4555; discussed later in Sect. \ref{sect:new_discovery}). \par 
Signs of interactions or merging have been previously published in all of these galaxies in group environments except for NGC~0383 (but see Sect. \ref{sect:results}). For example, the disturbed X-ray features in NGC~6861 also seen in SAUNAS (Fig. \ref{fig:categories}) have been previously linked to its merging with NGC~6868 in the Telescopium group \citep{machacek+2010apj711_1316}. Referring to the galaxies shown in  Fig. \ref{fig:small_rpeak}, similar arguments have been made in the NGC~1407 group \citep[see][]{ su+2014apj786_152} and NGC~4782 \citep[see][]{machacek+2008apj674_142}.

In the case of NGC~4261 (2nd row in Fig. \ref{fig:small_rpeak}), on the one hand the ``X''-shaped, X-ray features seen on the lower region the galaxy near \rpeak{} is thought to have been channeled by the large, powerful radio jets due to AGN activity \citep[][]{osullivan+2011mnras416_2916, temi+2022apj928_150}. On the other hand, the lack of diffuse X-ray emission in the upper part of the galaxy where very large optically detected tidal features have been found \citep[see][]{bilek+2020mnras498_2138, ebrova+2021aap650_50} is potentially a signature of the merger activity causing the outer X-ray halo shape asymmetry. \par 

The above findings suggest that mergers, not AGN activity, is the main process leading to the observed larger scale asymmetry in the more truncated hot halos. This interpretation is also supported by the ETGs in our sample where \rpeak $> R^{ETG}_{\rm edge}$ and whose X-ray halos are not significantly asymmetric. AGN feedback has in fact been reported in IC~1262 \citep[see the diamond shape of the halo within \rpeak{} in Appendix Fig. \ref{fig:no_asym} and in][]{pandge+2019apj870_62} as well as in NGC~1550 which is also now more relaxed after a past merger \citep[see Fig. \ref{fig:categories} and][]{kolokythas+2020mnras496_1471}. In contrast, evidence for neither recent AGN or merger activity have been found in NGC~4325, a galaxy with low asymmetry \citep[Appendix Fig. \ref{fig:no_asym} and][]{lagana+2015aap573_66}.  While the sample studied in this work lacks more ETGs with no significant asymmetry, the above findings suggest that the small \rpeak{} $<< R^{ETG}_{\rm edge}$ of these highly asymmetric and truncated  galaxies must have originated from a more recent external interaction or merger.  \par 

Interestingly, the conclusion that the location of \rpeak{} is more strongly influenced by an external process is also supported by the case of the cluster galaxy in our sample, NGC4374 or M84 which is located in Virgo (lower middle panel in Fig. \ref{fig:small_rpeak}). The peculiar ``H'' shape reported in \citet{bambic+2023mnras522_4374} and also detected in SAUNAS, is a consequence of an AGN jet. The southern tail is a signature of gas removal due to ram pressure in the high density environment of Virgo \citep{randall+2008apj688_208, shin+2013mnras428_804}. The authors show that there is no evidence for a temperature increase in the central regions of the halo structure caused by M84's super massive black hole. This result further supports the interpretation that the more truncated \rpeak{} must have been caused by an external event even in the case where environmental ram pressure processes are ongoing. \par

As a caveat to this interpretation, the full sample analyzed lacks galaxies with \rpeak{} $>> R^{ETG}_{\rm edge}$. The average stellar size of an ETG with $M_{\star}$ $\sim 10^{11}\,M_{\odot}$ is $\sim$ 30\,kpc. In future work it would be interesting to study a larger sample of ETGs with \rpeak{} $> 30$\,kpc and as a function of both merger and feedback/AGN stage similar to the work by \citet{smith+2019aj158_169}.

\subsection{Physical significance of the correlation between \rpeak{} and X-ray luminosity $L_X$}
\label{sect:correlation}

As discussed in \citet{kim+2020mnras492_2095}, the region $r$ of the X-ray halo in between the inner dip $R^X_{\rm break}$ and outer X-ray peak temperature radius \rpeak{}, i.e. $R^X_{\rm break} < r <$ \rpeak{} in the X-ray temperature profile, defines the cool core. It is interesting to speculate if the strong correlation between \rpeak{} and X-ray luminosity $L_X$ suggests that the X-ray halo region enclosed by the boundary of the cool core (\rpeak{}) reaches self-similarity, regardless of the impact of feedback or external interactions on \rpeak{} and $L_X$. \par 
The concept of self-similarity in X-ray halos is the idea that the properties of halos should scale with mass and redshift alone. Self-similarity assumes that the host galaxy formed via gravitational collapse and that the virial theorem is obeyed \citep[][]{kaiser1986mnras222_323, bohringer+2012aap539_120}. Typically, deviations from the self-similar predictions in X-ray scaling relations are associated with non-gravitational processes such as feedback and mergers \citep[see also the reviews by][]{nardini+2022incollection_90, bogdan+2022incollection_111}. Under the assumptions outlined above, the potential link between the \rpeak{}-- $L_X$ relation and self-similarity is discussed below. \par 

Following \citet[][their Eqs. 9 and 20;]{lovisari+2022incollection_65}, if the X-ray halo is isothermal with temperature $T_X$ and is in virial equilibrium, the virial theorem implies that:

\begin{equation}
\label{eq:virial}
    T_X \propto M/R
\end{equation}

where $M$ is the mass of the galaxy and $R$ is the radius of a sphere that encloses $M$. For self-similar halos, the density of the plasma is  $\rho \propto M/R^3$ and $L_X \propto T_X^{2}$. Combing these relations with Eq. \ref{eq:virial}, $T_X$ and $R$ can be expressed as:

\begin{eqnarray}
    T_X \propto R^2 \propto L_X^{0.5} \\
    R \propto L_X^{0.25} 
\label{eq:rt}  
\end{eqnarray}

If the $L_{X, soft}-T_X$ relation was used where $L_{X, soft}$ is the X-ray soft-band luminosity, then the dependence with $T_X$ in the self-similar model is $L_{X, soft} \propto T_X^{1.5}$ \citep[see Table 1 in][]{lovisari+2022incollection_65}. Eq. \ref{eq:rt} then changes to $R \propto L_X^{0.33}$. \par 
Using a moderate sample of ETGs, we have found that the slope $\beta$ of the \rpeak{}--$L_X$ relation is compatible with the above predictions from self-similarity with $\beta=0.27\pm0.09$. The comparable slope implies that feedback or mergers have a minor impact on the \rpeak{}--$L_X$ relation. At first glance, this finding appears to contradict the strong correlation between the ratio \rpeak{}/$R^S_{\rm edge }$ and X-ray halo shape asymmetry (Fig. \ref{fig:rpeak_corr}) because asymmetry can be caused by feedback and mergers. However, the large scatter in the \rpeak{}--$L_X$ relation $\sim 0.14$\,dex is a hint that feedback and mergers do in fact have a non-negligible impact on the distribution of galaxies in the \rpeak{}--$L_X$ plane (and thus on cool cores). \par
This interpretation is based on a previous discovery considering the neutral atomic hydrogen distribution (\HI) in galaxies. The tight correlation between the size of the \HI{} disk $D_{\HI{}}$, the diameter defined by the 1\,$M_{\odot}$/pc$^2$ contour, and total \HI{} mass of galaxies has been shown to apply to galaxies in all environments and evolutionary stages \citep[][]{wang+2016mnras460_2143, rajohnson+2022mnras512_2697}. This characteristic is found regardless of the clear \HI{} asymmetric features observed in individual galaxies due to feedback, galaxy interactions or environmental processes. \par 
A study using hydrodynamical simulations have found that the \HI{} component in galaxies re-distributes due to these processes in such a way that the galaxy still follows the $D_{\HI}-M_{\HI}$ relation \citep[see Fig. 9 in][]{stevens+2019mnras490_96}. This behavior is reflected in the scatter of the $D_{\HI}-M_{\HI}$ relation which remains small at $0.07$\,dex across galaxy environments. This scatter is a factor of two smaller than the \rpeak{}--$L_X$ shown here. \par 

A similar effect could potentially be happening in the \rpeak{}--$L_X$ relation. But the comparatively larger scatter in this relation $\sim 0.14$\,dex suggests that the dynamics by which X-ray halos reach equilibrium after an episode of feedback or merger may not be the same as in the \HI{} component. It would be interesting to search for deviations from this relation using a larger sample of ETGs in future studies to investigate this issue further. 

\subsection{Analysis of new halo features with SAUNAS/Chandra and DECaLS}
\label{sect:new_discovery}

The following discussion on the new discoveries found using SAUNAS/\emph{Chandra} and DECaLS refer to Fig. \ref{fig:stream_NGC0383} for galaxy NGC~0383 (Sect. \ref{subsec:ngc383}) and Fig. \ref{fig:ngc4555_1600} for galaxies NGC~1600 (Sect. \ref{subsec:ngc1600}) and 4555 (Sect. \ref{subsec:ngc4555}).

\subsubsection{NGC~0383}
\label{subsec:ngc383}

NGC~0383 is radio galaxy located in a highly complex environment as part of the Arp 331 chain made up of E and S0 galaxies \citep[][]{komossa+1999aap344_755, trussoni+1997aap327_27, heesen+2018mnras474_5049}. The upper panel of Fig. \ref{fig:stream_NGC0383} shows the main X-ray emitting galaxies in this chain. The lower panels visualize the intra-group light (IGL) in $g$-band surface brightness $\mu_g$, defined where $\mu_g > 26$~ \magarc\ and plotted as the dark regions connected to NGC0383. \par 

If the galaxies and region of the chain is bounded by the full segmentation map shown in the right panel, the average $g-r$ color of the chain is $\sim$ 0.86, while that of the IGL alone is $\sim$ 0.67. The fraction of IGL is $\sim$  2.7\% in the g-band which is consistent with that expected for small galaxy halos $\sim 10^{11}\,M_{\odot}$ \citep{purcell+2007apj666_20}. These fractions are much lower than those reported for intra-cluster light at similar redshift where the fractions can be $\sim 10\%$ or higher \citep[see Fig. 2 in][]{montes2022nat6_308}. Given that the depth of DECaLS is $\sim$2 mag shallower than the current deep surveys typically used to study IGL, the values estimated here must be considered as lower limits. \par 

Despite the shallower depth of DECaLS,  we detect a faint tidal feature (yellow box in the upper middle and right panels of Fig. \ref{fig:ngc4555_1600}). Detecting this full tidal feature as a single connected source is challenging, and demonstrates the need for ad hoc methods in order to fully analyze this group environment. Nevertheless, the discovery of this large tidal feature, potentially larger than the stellar size of a $M_{\star} \sim 10^{11.4}\,M_{\odot}$ ETG $\sim 45$\,kpc, indicates a recent interaction with one of its members. While it is unclear from our data alone if the perturber is the closest galaxy to NGC0383, NGC0386 \citep[see][]{blandford+1978mnras185_527}, an interaction could have triggered the AGN of NGC~0383 to emit its observed powerful jet \citep[e.g.][]{trussoni+1997aap327_27}. This scenario could explain the large-scale asymmetry of the X-ray halo (first panel in Fig. \ref{fig:small_rpeak}). Additionally, the detection of this stream suggests that the formation of this intra-group environment is an ongoing process \citep[see][and references therein]{montes2022nat6_308}. 
Deeper imaging can be used to confirm the association of these detections with NGC~0383 and assess the impact of depth, PSF and IGL definitions on the observed fraction \citep[see also][]{martinezlombilla+2023mnras518_1195, brough+2024mnras528_771}. \par 

We also point out that the X-ray halo of the Arp 331 Chain potentially extends out to much larger radial distances than the \emph{Chandra} maps analyzed here \citep[see the figures shown in][]{trussoni+1997aap327_27, komossa+1999aap344_755, heesen+2018mnras474_5049}. It would be interesting to investigate if the IGL shown here closely follows the distribution of the total mass of the Arp 331 chain as in clusters \citep[][and references therein]{montes+2019mnras482_2838, alonsoasensio+2020mnras494_1859, butler+2025arXiv2504.03518}. In a future study we will investigate how FoV and resolution impacts low surface brightness asymmetry analysis for such massive halos.

\subsubsection{NGC~1600}
\label{subsec:ngc1600}

As labeled in the upper panels of Fig. \ref{fig:ngc4555_1600}, NGC~1600 has two neighboring galaxies NGC~1601 and 1603. The \citet{devaucouleurs+1991book} catalog classified the former as an edge-on ``S0?'' galaxy and the latter as a potential early-type ``E?''. Both galaxies are also listed in the Chandra Source Catalog \citep{2024yCat.9070....0E} and detected in SAUNAS. Although NGC~1600 hosts a faint ultra-massive black hole \citep{thomas+2016nat532_340}, the galaxy has a cool core \citep{kim+2020mnras492_2095} and \citet{runge+2021mnras502_5487} show that even the dynamics of the hot plasma within 3\,kpc from the center is not dictated by this black hole. This result implies that the shape of the larger scale asymmetric X-ray halo detected in SAUNAS must have originated from an external event. \par
In fact, the large wedged-shape of the halo, spanning $\sim 25$\,kpc on each side, and the lack of X-ray emission on the top right of the galaxy is reminiscent of a cold front \citep[see Fig. 19 in][]{ascasibar+2006apj650_102}. This shape is similar to that of the merging galaxy NGC~6861 mentioned earlier where a cold front edge has been detected in \citet[][]{machacek+2010apj711_1316} and in SAUNAS.\par

\citet{kim+2019apj241_36} did not report this front feature in their X-ray maps and argue that NGC~1603 is likely too small to have created the front structure. Using the DECaLs g- and r-band images, we estimate that the stellar mass of NGC~1603 is at least $\sim 3.3\times10^{10}\,M_{\odot}$\footnote{Following \citet{roediger+2015mnras452_3209}, we use the global $g-r$ color of the galaxy to estimate the stellar mass-to-light ratio, based on the \citet{bruzual+2003mnras344_1000} stellar population synthesis model. The segmentation map is used to select the region belonging to the galaxy. Given that there is significant overlap with its host, and the galaxy has likely lost stars by tidal interactions, this stellar mass value must be considered as a lower limit.} which is massive enough to cause the cold front according to simulations \citep{ascasibar+2006apj650_102}. Given that we have additionally found tidal features in star light using DECaLS, we can confirm that NGC~1603 is disturbed in its outer region and merging with its host.   \par 

The above analysis is also in contradiction with \citet{sivakoff+2004apj617_262} who (1) interpret the X-ray tail emanating from NGC~1603 as purely a sign of ram pressure and (2) report an excess of plasma in the North-East of the galaxy. In contrast, SAUNAS detects an excess in the South which is highly asymmetric. Combining the new detections in  X-ray and optical, we propose that the X-ray tail emanating from NGC~1603 may not only stem from ram pressure but additionally due to the tidal removal of its gas and stars in the merging process with its host NGC~1600. \par

\subsubsection{NGC~4555}
\label{subsec:ngc4555}

In the case of NGC4555, previous work report that this galaxy is in an isolated environment and hosts a smooth X-ray halo \citep[e.g.][]{osullivan+2004mnras354_935, goulding+2016apj826_167, kim+2019apj241_36}. While SAUNAS/\emph{Chandra} detects the halo to be as extended as that reported in e.g. \citet[$\sim$ 60\,kpc, although not as a contiguous source as in that work;][]{osullivan+2004mnras354_935}, the diffuse X-ray map presented in Fig. \ref{fig:ngc4555_1600} also reveals several asymmetric features, one in the direction of its nearest neighbors WISEA J123541.33+262856.7 (RA=188.92 deg,
DE=26.48 deg) and
WISEA J123535.89+263400.0 (RA=188.90, DE=26.57) both of which are also at a redshift of 0.022\footnote{\protect\url{https://ned.ipac.caltech.edu}} \citep{helou+1991inproceedings_89}. These galaxies are detected in the SAUNAS X-ray map. \par 
While these galaxies are more than a factor of ten fainter than NGC~4555 (DECaLS r-band apparent magnitudes of 15.05\,mag and 15.24\,mag, respectively), at this distance we estimate their stellar masses to be $\sim 2.3\times 10^{10}\,M_{\odot}$ and $1.7\times10^{10}\,M_{\odot}$, respectively. Given the absence of any spiral sub-structures in them, they appear to have early-type morphology. \par
As highlighted using segmentation maps in Fig. \ref{fig:ngc4555_1600}, diffuse light ``bridges'' in the optical is additionally detected in the outer regions of these neighbors in the direction of NGC~4555. These results suggest that NGC~4555 is tidally interacting with  satellite galaxies WISEA J123541.33+262856.7 and WISEA J123535.89+263400.0 and these galaxies are massive enough to potentially disturb the X-ray halo of their host galaxy. However, compared to the previous galaxy NGC~1600 discussed above, the satellites of NGC~4555 do not appear to have any obvious X-ray tails in the direction of their host which can be used to confirm this scenario. \par 
It is also interesting to point out that the overall disturbed shape of NGC~4555's  X-ray halo is strikingly similar to NGC~741 shown earlier in Fig. \ref{fig:small_rpeak}. The X-ray structure of NGC~741 is thought to be due to a group satellite that recently passed right through the core of the group near NGC~741 \citep[see][]{schellenberger+2017apj845_84}. While the optical data suggests that several satellites appear to be on in-fall or merging with NGC~4555 such as those satellites analyzed here, it is unclear whether NGC~4555 suffered a similar merger history as NGC~0741. Deeper imaging could be used to search for X-ray tidal tails in the satellites of NGC~4555 in order to investigate the origin of this unusual X-ray halo shape in future work.

\section{Conclusions}

Key findings described in this study include: 
(1) Early-type galaxies with significant asymmetry in their X-ray halos have more truncated peak temperature radius with respect to their stellar size (\rpeak $<< R^{ETG}_{edge}$). This negative correlation between the ratio \rpeak{}/$R^{ETG}_{edge}$ and shape asymmetry $A_S$ is statistically significant (p-value $\sim 10^{-5}$). The detected X-ray asymmetries in the more truncated halos can be associated with a recent external event such as a galaxy interaction or merger. This result supports the hypothesis that the location of \rpeak{} is sensitive to these processes that can impact the shape of the X-ray halo. 

(2) The positive correlation between \rpeak{} and X-ray halo luminosity $L_X$ where \rpeak{}$ \sim L_X^{0.27}$ is statistically significant (p-value $\sim 10^{-6}$). Although the slope of the relation is compatible with the predictions for self-similar X-ray halos, the scatter of the relation is large $\sim 0.14$\,dex. This scatter potentially reflects the impact of feedback and galaxy interactions on the distribution of galaxies in the \rpeak{}--$L_X$ plane. Deviations from this relation can be examined with a larger sample of galaxies at different feedback and merger stages in future studies. 

(3) Newly discovered asymmetric features using SAUNAS and DECaLS imaging in Sect. \ref{sect:results} are presented for galaxies NGC~1600 and 4555. In the case of NGC~1600, these new detections confirm the tidal interactions with its satellites. These satellites have stellar masses of the order of $\sim 10^{10}\,M_{\odot}$ which make them massive enough to disturb the hot plasma in their host. \par 
In the case of NGC~4555, stellar tidal features are also detected which indicate a recent interaction with its $\sim 10^{10}\,M_{\odot}$ satellites. However, there are no obvious signs of X-ray disturbance in these satellites.  Moreover, the X-ray halo shape of NGC~4555 is highly unusual for a galaxy in a low dense environment. Further investigation is necessary to understand the origin of this unusual halo shape.

(4) NGC~0383 is in the complex environment of the Arp 331 chain. The galaxy also shows clear signs of disturbance in the X-ray halo at kpc scales. We additionally discovered a stellar tidal stream in the outer regions of NGC~0383, showing clear evidence of a recent  interaction in this group environment for the first time. This interaction could have triggered the AGN in NGC~0383 to emit its powerful radio jet. The detection of the stream also shows that the IGL of the chain is ongoing assembly. These results underscore the importance of truncations and deep, multi-wavelength observations in order to disentangle the processes responsible for the assembly of ETGs.

\begin{acknowledgments}
The authors thank the anonymous referee for their critical feedback which greatly improved the presentation of the results and clarity of this work. NC sincerely thanks Ewan O'Sullivan, Dong-Woo Kim and Giuseppina (Pepi) Fabbiano for several interesting discussions and feedback related to this work during her visit to the Harvard-Smithsonian Center for Astrophysics. 
N. C.'s research is supported by an appointment to the NASA Postdoctoral Program at the NASA Ames Research Center, administered by Oak Ridge Associated Universities under contract with NASA. Support for this work was provided in part by the National Aeronautics and Space Administration through Chandra Award \#24610329 issued by the Chandra X-ray Center, which is operated by the Smithsonian Astrophysical Observatory for and on behalf of the National Aeronautics Space Administration under contract NAS8-03060.

This research has made use of the NASA/IPAC Extragalactic Database (NED), which is funded by the National Aeronautics and Space Administration and operated by the California Institute of Technology.

This paper employs a list of Chandra datasets, obtained by the Chandra X-ray Observatory, contained in the Chandra Data Collection (CDC) ~\dataset[doi:10.25574/cdc.375]{https://doi.org/10.25574/cdc.375}, from the Chandra Source Catalog and software provided by the Chandra X-ray Center (CXC) in the application packages \texttt{CIAO} \citep{fruscione+2006inproceedings_62701V} and \texttt{Sherpa} \citep{freeman+2001inproceedings_76}. 

This work is based in part on observations made with the Spitzer Space Telescope, which was operated by the Jet Propulsion Laboratory, California Institute of Technology under a contract with NASA.

The Legacy Surveys consist of three individual and complementary projects: the Dark Energy Camera Legacy Survey (DECaLS; Proposal ID 2014B-0404; PIs: David Schlegel and Arjun Dey), the Beijing-Arizona Sky Survey (BASS; NOAO Prop. ID 2015A-0801; PIs: Zhou Xu and Xiaohui Fan), and the Mayall z-band Legacy Survey (MzLS; Prop. ID 2016A-0453; PI: Arjun Dey). DECaLS, BASS and MzLS together include data obtained, respectively, at the Blanco telescope, Cerro Tololo Inter-American Observatory, NSF’s NOIRLab; the Bok telescope, Steward Observatory, University of Arizona; and the Mayall telescope, Kitt Peak National Observatory, NOIRLab. Pipeline processing and analyses of the data were supported by NOIRLab and the Lawrence Berkeley National Laboratory (LBNL). The Legacy Surveys project is honored to be permitted to conduct astronomical research on Iolkam Du’ag (Kitt Peak), a mountain with particular significance to the Tohono O’odham Nation.

NOIRLab is operated by the Association of Universities for Research in Astronomy (AURA) under a cooperative agreement with the National Science Foundation. LBNL is managed by the Regents of the University of California under contract to the U.S. Department of Energy. \\

This project used data obtained with the Dark Energy Camera (DECam), which was constructed by the Dark Energy Survey (DES) collaboration. Funding for the DES Projects has been provided by the U.S. Department of Energy, the U.S. National Science Foundation, the Ministry of Science and Education of Spain, the Science and Technology Facilities Council of the United Kingdom, the Higher Education Funding Council for England, the National Center for Supercomputing Applications at the University of Illinois at Urbana-Champaign, the Kavli Institute of Cosmological Physics at the University of Chicago, Center for Cosmology and Astro-Particle Physics at the Ohio State University, the Mitchell Institute for Fundamental Physics and Astronomy at Texas A\&M University, Financiadora de Estudos e Projetos, Fundacao Carlos Chagas Filho de Amparo, Financiadora de Estudos e Projetos, Fundacao Carlos Chagas Filho de Amparo a Pesquisa do Estado do Rio de Janeiro, Conselho Nacional de Desenvolvimento Cientifico e Tecnologico and the Ministerio da Ciencia, Tecnologia e Inovacao, the Deutsche Forschungsgemeinschaft and the Collaborating Institutions in the Dark Energy Survey. The Collaborating Institutions are Argonne National Laboratory, the University of California at Santa Cruz, the University of Cambridge, Centro de Investigaciones Energeticas, Medioambientales y Tecnologicas-Madrid, the University of Chicago, University College London, the DES-Brazil Consortium, the University of Edinburgh, the Eidgenossische Technische Hochschule (ETH) Zurich, Fermi National Accelerator Laboratory, the University of Illinois at Urbana-Champaign, the Institut de Ciencies de l'Espai (IEEC/CSIC), the Institut de Fisica d’Altes Energies, Lawrence Berkeley National Laboratory, the Ludwig Maximilians Universitat Munchen and the associated Excellence Cluster Universe, the University of Michigan, NSF’s NOIRLab, the University of Nottingham, the Ohio State University, the University of Pennsylvania, the University of Portsmouth, SLAC National Accelerator Laboratory, Stanford University, the University of Sussex, and Texas A\&M University.
\\

BASS is a key project of the Telescope Access Program (TAP), which has been funded by the National Astronomical Observatories of China, the Chinese Academy of Sciences (the Strategic Priority Research Program ``The Emergence of Cosmological Structures'' Grant XDB09000000), and the Special Fund for Astronomy from the Ministry of Finance. The BASS is also supported by the External Cooperation Program of Chinese Academy of Sciences (Grant 114A11KYSB20160057), and Chinese National Natural Science Foundation (Grant 12120101003, 11433005). \\

The Legacy Survey team makes use of data products from the Near-Earth Object Wide-field Infrared Survey Explorer (NEOWISE), which is a project of the Jet Propulsion Laboratory/California Institute of Technology. NEOWISE is funded by the National Aeronautics and Space Administration.

The Legacy Surveys imaging of the DESI footprint is supported by the Director, Office of Science, Office of High Energy Physics of the U.S. Department of Energy under Contract No. DE-AC02-05CH1123, by the National Energy Research Scientific Computing Center, a DOE Office of Science User Facility under the same contract; and by the U.S. National Science Foundation, Division of Astronomical Sciences under Contract No. AST-0950945 to NOAO.

\vspace{5mm}
\facilities{Spitzer (IRAC), \emph{Chandra} (ACIS), Blanco (DECam)}


\software{\texttt{Astropy} \citep{collaboration+2018aj156_123, collaboration+2013aap558_33, collaboration+2022apj935_167}, \texttt{CIAO}, \texttt{LIRA} \citep{donath+2022inproceedings_98}\footnote{\texttt{pyLIRA:} \url{https://github.com/astrostat/pylira}}, \texttt{Matplotlib} \citep{hunter2007sci9_90}, 
\textit{Matrioska} \citep[mascara function;][]{borlaff+2018aap615_26},
\texttt{NumPy} \url{http://www.numpy.org/},
\texttt{VorBin} \citep{cappellari+2003mnras342_345}, \texttt{SAO Image DS9} \citep{softwareDS9},
\texttt{SciPy} \url{http://www.scipy.org/}, 
\texttt{Sourcerer} \citep{teeninga+2016MMTA1_100, haigh+2021aap645_107},
\texttt{SWarp} \citep{2010ascl.soft10068B}}

\end{acknowledgments}

\appendix 

\section{Sample selection and potential biases}
\label{app:table}

Table \ref{tab:sample} summarizes the sample of galaxies in our study. The original CGA sample is taken from \citet{kim+2020mnras492_2095} and we focus on the sub-sample of 30 galaxies  with identified \rpeak{} (see Sect. \ref{sect:data}). The number of galaxies with available imaging in the S4G and DECaLS surveys are also listed in the Table \ref{tab:sample}. 

\begin{table}[h]
    \centering
    \begin{tabular}{c|c}
       Sample  &  No. of Galaxies\\ \hline 
       CGA (\rpeak{}) & 60 (30) \\
       CGA + S4G & 18 (6) \\
       \rpeak{} + DECaLS & 24 (19) \\
       Final &  38 (25)  \\ \hline

    \end{tabular}
    \caption{Summary of samples used in this work. Values in brackets indicate the number of galaxies selected in the final sample with \rpeak{} measurements. See text for details.}
    \label{tab:sample}
\end{table}

Regarding any potential biases, the final sample selected lacks ETGs with very large \rpeak{} $>> R^{ETG}_{\rm edge}$ (see Fig. \ref{fig:edges_peak_radii} and Fig. \ref{fig:rpeak_corr}). If galaxies with large \rpeak{} can also have highly asymmetric X-ray halos, then this result could potentially weaken our main conclusions. This issue is explored in the following. \par

One of the main criteria used in the sample selection procedure is the removal of galaxies with high contamination from bright stars or overlapping sources in their group environments in DECaLS imaging. This criterion is required for the characterization of their stellar edges (see Sect. \ref{sect:data}).  The \rpeak{} of the five galaxies which were omitted in the final sample due to this contamination is also in the range 13-70\,kpc. Two of these galaxies have an \rpeak{}$ >> R^{ETG}_{\rm edge}$ for their $M_{\star}$, namely NGC~0507 and 6338. Below we summarize the environment and known X-ray asymmetries of these galaxies (if any) as a simple test of the robustness of our conclusions despite the lack of galaxies with \rpeak{}$ >> R^{ETG}_{\rm edge}$ in the final sample. \par
NGC0507 is in a rich environment and considered the  brightest cluster galaxy in its group \citep{kim+2020mnras492_2095}. The galaxy is known to have some X-ray asymmetry which is caused by an interaction with the neighboring galaxy NGC~0499 \citep{brienza+2022aap661_92}. By visual comparison of the surface brightness maps shown in \citet[see their Fig. 4;][]{brienza+2022aap661_92} with the final sample analyzed in this work, the X-ray asymmetry of NGC~0507 is similar to the asymmetry seen in NGC~5846 where $A_S = 0.46\pm0.01$  which is presented in Appendix Fig. \ref{fig:def_asym}. \par

On the other hand, NGC~6338 is in a less crowded group and has an X-ray morphology similar to IC~1262 (Appendix Fig. \ref{fig:no_asym}, $A_S = 0.35\pm0.01$) but with inner filaments at much smaller scale $< 6$\,kpc due to an AGN feedback \citep[see Fig. 2 in][]{pandge+2012mnras421_808}. The system is to merge with galaxy MCG +10-24-117 but they are still separated and have not passed through the core \citep[see the more recent work by][]{schellenberger+2023apj948_101}. \par 
This investigation reveals that the NGC~0507 and 6338 galaxies with large \rpeak{} have minor asymmetries in the outer regions of their X-ray halo compared to the more extreme cases found with smaller \rpeak{} (see Fig. \ref{fig:stream_NGC0383}-\ref{fig:small_rpeak}). Therefore, the exclusion of these galaxies with large \rpeak{} in the final sample due to the contamination of the optical data does not change our main conclusions. While beyond the scope of this study, in future work it would be interesting to study \rpeak{} as a function of both merger and feedback/AGN stage using a larger sample, similar to the work by \citet{smith+2019aj158_169}. \par

\section{Uncertainties in $R^S_{\rm edge}$ and $R^X_{\rm peak}$}
\label{app:upper_lims}

\textbf{$R^S_{\rm edge}$}: The main source of uncertainty in $R^S_{\rm edge}$ is from the sky background estimation in the images which can impact the shape of the radial profile in the outer regions as well as the uncertainty in visually identifying $R^S_{\rm edge}$ repeatedly by expert observers. Briefly, the average uncertainty in $R^S_{
\rm edge}$ due to the sky background estimation is computed by moving the radial profiles by a value randomly chosen from a normal distribution with its standard deviation set by the uncertainty of the image background. This process is repeated 100 times for each galaxy. The total uncertainty for each galaxy due to the sky background is computed by adding in quadrature the uncertainty in both image bands used to derive the color profile, resulting in a combined uncertainty of 0.13\,kpc on average. The typical uncertainty considering repeated identification of the edge feature by expert observers is $0.04$\,dex \citep[see][]{chamba+2022aap667_87}, corresponding to $\sim 1.1$\,kpc on average. Both the uncertainties due to the image background for each galaxy and the 0.04\,dex error from repeated identifications are used in the $R^S_{\rm edge}$ measurements shown in Fig. \ref{fig:edges_peak_radii}.

\textbf{$R^X_{\rm peak}$}: In the case of \rpeak{}, the radii values are initially taken from the measurements made by \citet{kim+2020mnras492_2095}. The X-ray peak temperature radius listed in that work corresponds to the radial location of the peak temperature in the best-fit, 2D-projected temperature model to the observed data points in the temperature profile (see their Sect. 2.2 and Fig. 1 which shows the different temperature profile types studied in that work). As we are interested in comparing the location of \rpeak{} with the X-ray halo shape asymmetry, we must ensure that these \rpeak{} values are within the binary halo detection mask where asymmetry is measured (Sect. \ref{sect:saunas} and Fig. \ref{fig:masking_connectivity}). This expectation is reasonable given that the \rpeak{} values were also measured using publicly available \emph{Chandra} archival data in \citet{kim+2020mnras492_2095}. \par 

Given that the halo binary detection mask is derived by thresholding the SAUNAS SNR map at 3$\sigma$, we require the \rpeak{} radius from the \citet{kim+2020mnras492_2095} analysis to be at least within the 1$\sigma$ contour in the SAUNAS SNR map. We select this threshold because the data products used in that work (from the Chandra Galaxy Atlas or CGA) have not been de-convolved from the effect of the PSF \citep{kim+2019apj241_36}. The PSF could make sources appear more extended in the CGA data products \citep[see][where the PSF de-convolution procedure is presented in SAUNAS]{borlaff+2024apj967_169}. Therefore, a low threshold such as 1$\sigma$ in SAUNAS allows us to be as inclusive as possible to the \rpeak{} values determined by \citet{kim+2020mnras492_2095} who used CGA data products.   \par

\begin{figure*}
\centering 
 \includegraphics[width=0.498\linewidth]{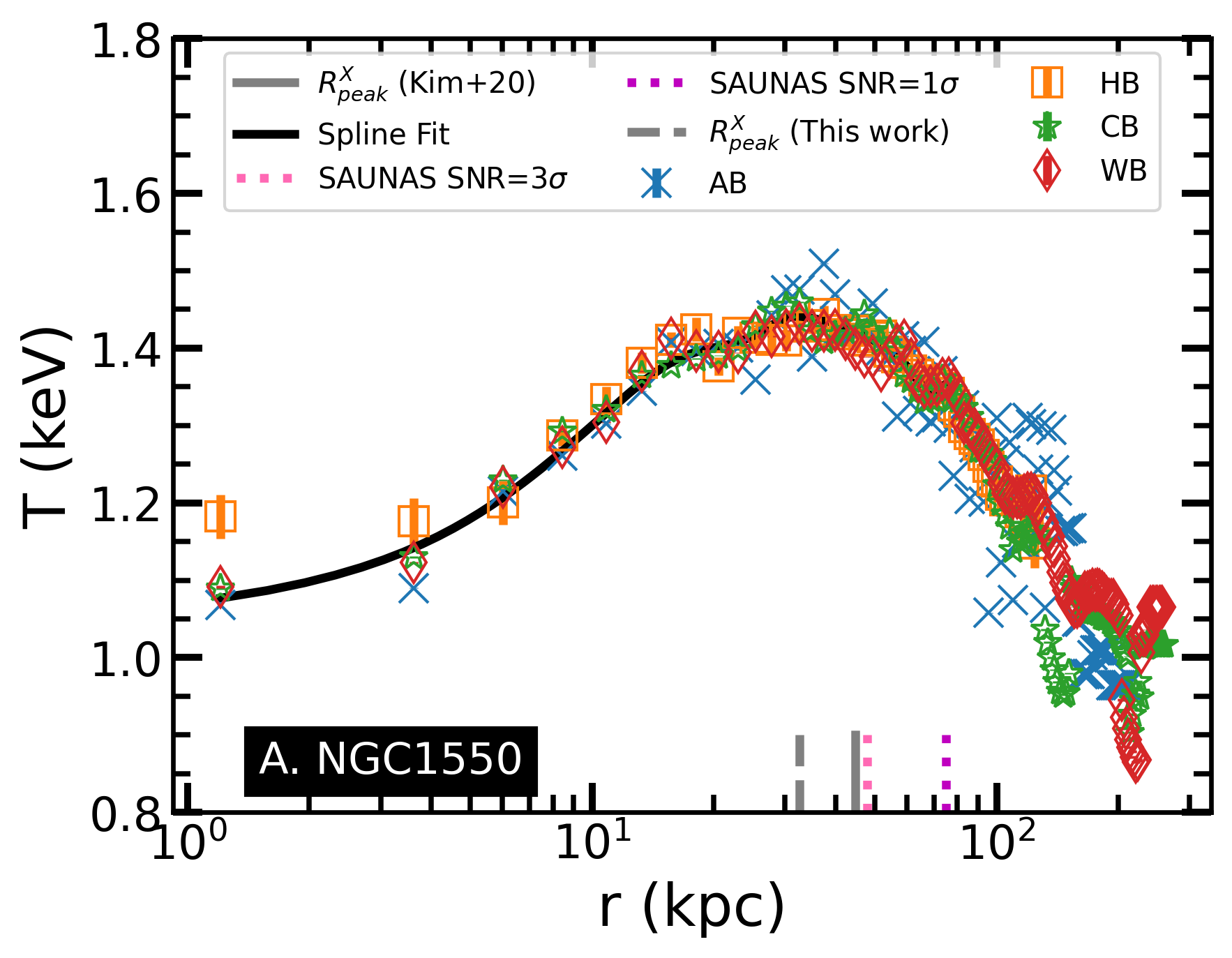} 
\includegraphics[width=0.47\linewidth]{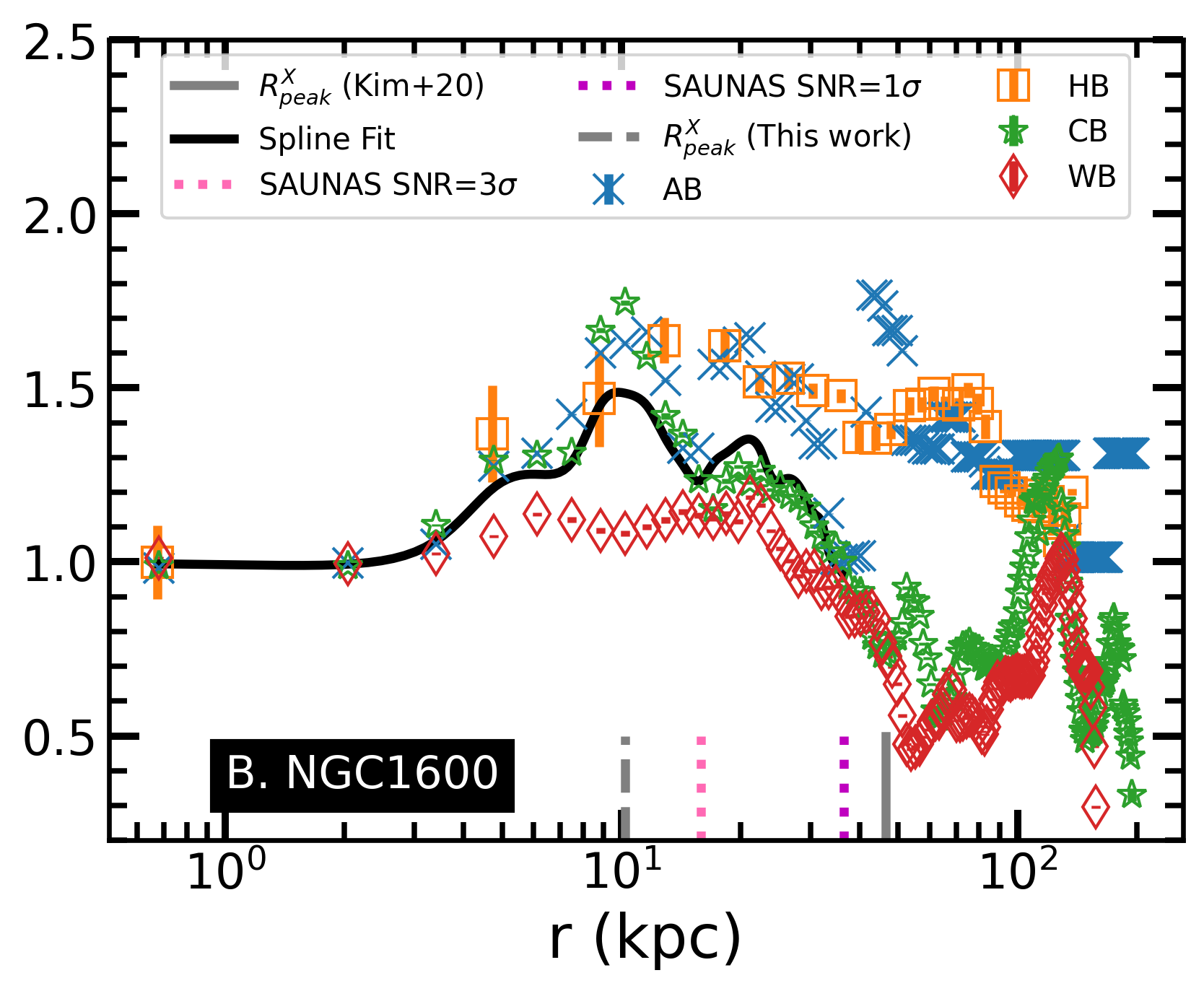} 
 
 \caption{Temperature profiles of NGC~1550 (panel A, left) and NGC~1600 (panel B, right) using the publicly available CGA temperature maps as illustrative examples. The X-ray halo shape asymmetry of NGC~1550 is low (0.36$\pm 0.01$, see Fig.  \ref{fig:masking_connectivity}) while NGC~1600 is high (0.92$\pm 0.03$, see Fig.  \ref{fig:ngc4555_1600}). The labeled temperature profiles were derived using the four adaptively binned temperature maps created by the CGA team: AB (annular binning), WB (weighted binning), CB (contour binning) and HB (hybrid binning). We refer the reader to \citet{kim+2019apj241_36, kim+2020mnras492_2095} for the details on those different binning methods. The profiles shown here were derived by azimuthally averaging the temperature maps using circular annuli, similar to the method used in \citet{kim+2020mnras492_2095}. The radius locations of the X-ray temperature peak radius from \citet{kim+2020mnras492_2095} (grey), this work (grey dashed), the radial location of the 3$\sigma$ (pink dotted) and 1$\sigma$ threshold in the SAUNAS SNR map (purple dotted) are indicated as large ticks in the x-axis and labeled in the legend. The radius locations of the 3$\sigma$ and 1$\sigma$  SAUNAS SNR is the average radial location of the aforementioned SNR contour in the SAUNAS map (see Fig. \ref{fig:masking_connectivity} for an example). We plot a spline fit to the data which is estimated using the mean of the four radial profiles at each radius bin. This function is shown only up to the SAUNAS 1$\sigma$ threshold radius and used to determine at which radius the maximum peak temperature occurs. The \citet{kim+2020mnras492_2095} X-ray temperature peak radius is within the X-ray halo region detected by SAUNAS at both 1 and 3 $\sigma$ for NGC~1550 (left) but not for NGC~1600 (right). For this reason, the derived \rpeak{} in this work for NGC~1600 is considered as a lower limit. See the text for more discussion.} 
 \label{fig:ngc1550_profile}
\end{figure*}

Ten galaxies in the \rpeak{} sample are found to miss this criterion -- in other words, their \rpeak{} is beyond the location where the SAUNAS SNR is 1$\sigma$. In these cases, the temperature profiles using all the maps provided in the CGA website are re-derived and the peak is located as the radius bin with the highest temperature. The methods used for this task are illustrated in Fig. \ref{fig:ngc1550_profile}. The figure displays the temperature profiles derived using CGA data products for two galaxies in our sample: one with a low X-ray halo shape asymmetry (NGC~1500) and another with a highly asymmetric X-ray halo (NGC~1600). \par

The larger scatter in the temperature profiles from more asymmetric X-ray halos was also mentioned in \citet{kim+2020mnras492_2095}. In the paragraph before Section 3.1 in \citet{kim+2020mnras492_2095}, the authors write: ``Non-spherically symmetric gas distributions manifest in the projected temperature profiles as vertical (temperature) scatter, due to the range of gas temperatures at the same radii. In this work, we intend to determine the shape of the global temperature profile and use the azimuthally averaged radial profiles. The azimuthal variation (e.g. analysing different pie sectors) will be addressed in future work.'' Given that \rpeak{} was derived in \citet{kim+2020mnras492_2095} using azimuthally averaged profiles (and not using pie sectors or other shapes), we follow this same approach to derive temperature profiles. 
The data and methods used for this task are described as follows.
\par 

The publicly available temperature maps (hereafter dubbed ``T-map'') are taken from the CGA \citep{kim+2019apj241_36,kim+2020mnras492_2095}\footnote{\protect\url{https://cxc.cfa.harvard.edu/GalaxyAtlas/v1/cga_main.html}}. These maps were created by extracting spectra from adaptively binned \emph{Chandra} data such that a certain SNR is reached for each bin. Four different adaptive binning techniques have been used in that study: AB (annular binning), WB (weighted binning), CB (contour binning) and HB (hybrid binning). We refer the reader to \citet{kim+2019apj241_36} for details on these different binning procedures. In the publicly available maps, the SNR for each bin was set to 20. \par 
The radial temperature profiles are derived for this sub-sample of galaxies using all these T-maps via azimuthal averaging. This technique was the same used to derive the temperature profiles shown in \citet{kim+2020mnras492_2095} and we use comparable step sizes for extracting the profiles (5 arcsec). \par 

The mean of the four radial profiles from the different binning methods is then used to create a smoothing spline function \footnote{\protect\url{https://docs.scipy.org/doc/scipy/reference/generated/scipy.interpolate.make_smoothing_spline.html}}. This spline fit allows us to visualize the overall, average shape of the temperature profile. The function is cut off at the SAUNAS 1$\sigma$ threshold radius and used to determine at which radius the maximum peak temperature occurs. The radius location of the 1$\sigma$ SAUNAS SNR is taken as the average radial location of the 1$\sigma$ contour in the SAUNAS map.\par

Following this approach, the uncertainty in \rpeak{} for all CGA galaxies is reported as the size of the radius bin where the highest temperature is found. The \rpeak{} located for these ten galaxies are considered lower limits. In other words, the cataloged values by \citet{kim+2020mnras492_2095} for these cases are considered as upper limits. 

Further justification for re-deriving \rpeak{} radii for galaxies where the \citet{kim+2020mnras492_2095} values are located beyond SAUNAS 1$\sigma$ SNR is made by considering a few individual cases. First, for galaxies NGC~1550 and NGC~4636, this radial profile method described above for identifying \rpeak{}  is consistent with the values reported in \citet{islam+2021apj256_22} who used XMM data. \par 
The \rpeak{} value derived for NGC~1550 is 32.6$\pm2.4$\,kpc which is within the range of 30-50\,kpc reported by \citet{islam+2021apj256_22} and lower than the cataloged value of 44.7\,kpc in \citet{kim+2020mnras492_2095}. All these values are well within the SAUNAS 3-sigma detections and the use of the cataloged value is consistent with our criteria. \par 
In the case of NGC~4636, the value derived is 13.2$\pm 0.5$\,kpc. The cataloged value of \rpeak{} by \citet{kim+2020mnras492_2095} is larger, at 25.4\,kpc which is beyond SAUNAS 1-sigma detections. As the peak radius reported in \citet{islam+2021apj256_22} is $\sim$ 15\,kpc, we use our re-derived value of 13.2\,kpc. Therefore, at least for these two cases, the above arguments verify the use of the radial temperature profiles to re-derive \rpeak{} in cases where the \citet{kim+2020mnras492_2095} values are identified as upper limits based on the SAUNAS detections. 

Second, if only the \citet{kim+2020mnras492_2095} values were used, then two main changes occur in the distribution of their \rpeak{} with respect to the $R^{ETG}_{\rm edge}$ relation. (1) Five additional galaxies populate the regime where \rpeak{} $> R^{ETG}_{\rm edge}$, namely NGC~4374 and NGC~4782 (from Fig. \ref{fig:small_rpeak}) and
NGC~4636, NGC~5044 and NGC~7619 (see Fig. \ref{fig:no_asym} and\ref{fig:def_asym}), bringing the total to nine instead of four galaxies with \rpeak{} $> R^{ETG}_{\rm edge}$ (or 17 galaxies with \rpeak{} $< R^{ETG}_{\rm edge}$). (2) The number of galaxies considered to have very small \rpeak{} $<< R^{ETG}_{\rm edge}$ is significantly reduced  from twelve to four galaxies: all the asymmetric galaxies $A_S > 0.4$ shown in Fig. \ref{fig:categories}, \ref{fig:ngc4555_1600} and \ref{fig:small_rpeak} except for NGC~1132, NGC~4261, NGC~4555 and NGC~6861 would no longer be populated below the shaded 3$\sigma$ $R^{ETG}_{\rm edge}$ relation. \par 
In the case of NGC~4636,  the use of our re-derived \rpeak{} value was already justified above. A single exception to this discussion is the case of NGC~5044 (which is also discussed in Appendix \ref{app:a_s_uncert}). Even though the shape asymmetry of NGC~5044's X-ray halo is low ($A_S = 0.29\pm0.1$; as noted in Sect. \ref{sect:results}), the \rpeak{} value of 26\,kpc positions this galaxy below the average $R^{ETG}_{\rm edge}$ relation (although still within the scatter of this relation). The \rpeak{} values reported in the literature are much larger: 53.9\,kpc by \citet{kim+2020mnras492_2095} and even larger $\sim$ 73\,kpc from the maps analyzed in \citet{david+2011apj728_162} and \citet{osullivan+2014mnras437_730}. These literature values would place NGC~5044 above the size--mass plane and would support our conclusions that galaxies with low asymmetry have larger \rpeak{}. The above findings confirm that even if the \citet{kim+2020mnras492_2095} cataloged values were used instead of our re-derived lower limits, the overall finding of this study that the majority of ETGs with asymmetry in their X-ray halos are also those with small \rpeak{} $ < R^{ETG}_{\rm edge}$ remains unchanged.  \par

Finally, by examining the rest of the sample mentioned above, we emphasize that all these galaxies except NGC~5044 have significant asymmetry according to $A_S > 0.6$. As \citet{kim+2020mnras492_2095} report \rpeak{} values which are located beyond the SAUNAS detections shown in Figs. \ref{fig:small_rpeak}--\ref{fig:def_asym}, these findings potentially indicate the difficulty in modeling the temperature distribution in the outer halos of highly asymmetric galaxies using radial profiles. While beyond the scope of this study to re-model the temperature distribution of the upper limit sub-sample, we emphasize that removing these galaxies from our analysis only reduces the sample size and does not change our main finding that the majority of galaxies with some or significant asymmetry in their X-ray halo have \rpeak{} $< R^{ETG}_{\rm edge}$.

\section{The uncertainty in the binary mask and $A_S$}
\label{app:a_s_uncert}

\begin{figure}
    \centering
    \includegraphics[width=0.32\linewidth]{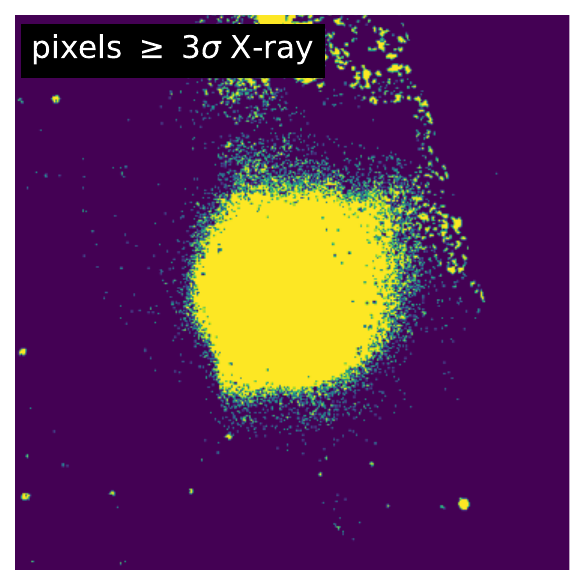}
    \includegraphics[width=0.32\linewidth]{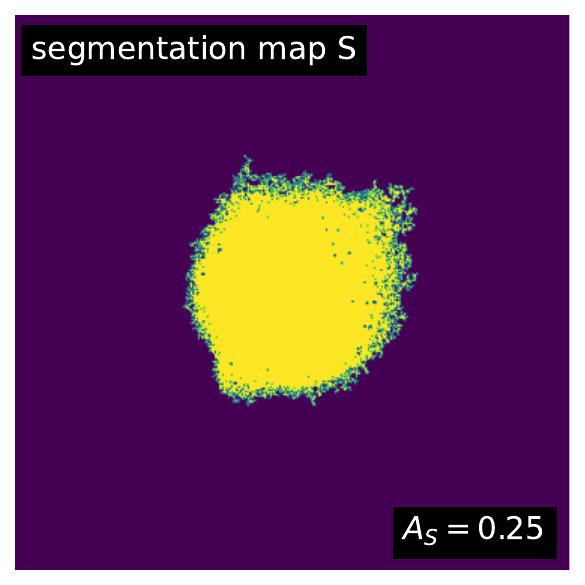} \\
    \includegraphics[width=0.32\linewidth]{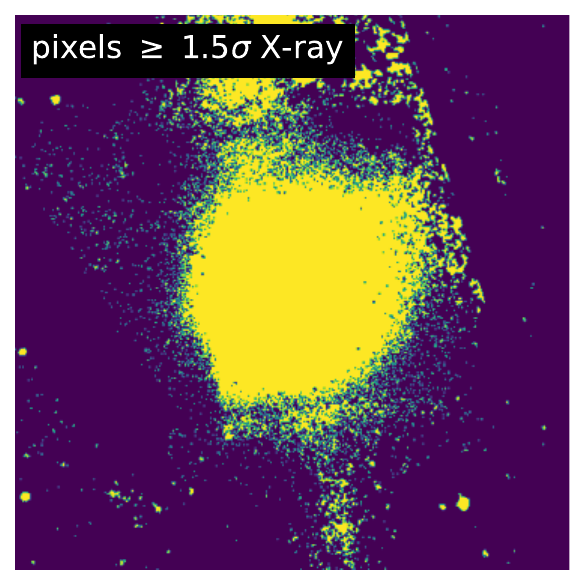}
    \includegraphics[width=0.32\linewidth]{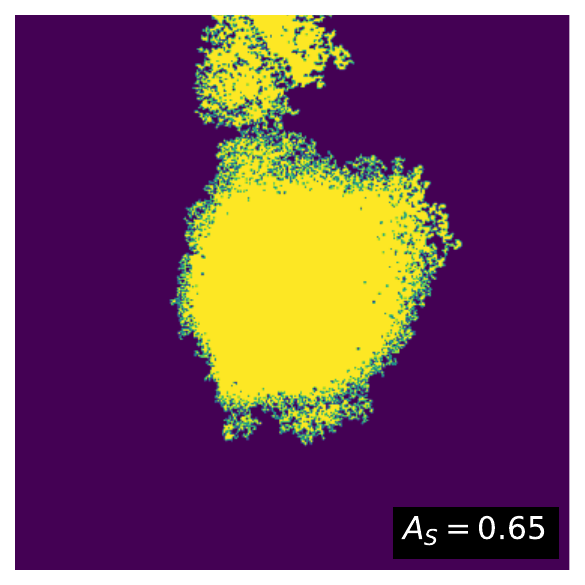}
    
    \caption{Segmentation map creation for NGC~5044. \textit{\textbf{Upper:}} Although the SAUNAS/\emph{Chandra} 3$\sigma$ map consists of significant pixels in the upper right of the galaxy (left), those pixels are not part of connected component of the galaxy according to the 8-connectivity operator (right). This map results in the lower $A_S$ value of 0.25 indicating no significant X-ray halo asymmetry. \textit{\textbf{Lower}}: If the significance level from SAUNAS is reduced by half to 1.5$\sigma$, then more pixels are connected to the main galaxy (right) and the $A_S$ value is significantly higher 0.65. This map would indicate that the X-ray halo of NGC~5044 is highly asymmetric. The figure shows that the estimation of $A_S$ is highly dependent on the input segmentation map.}
    \label{fig:ngc5044}
\end{figure}

Taking NGC~5044 as an example, in this section we demonstrate how the X-ray halo shape asymmetry parameter $A_S$ depends on the input binary detection mask or segmentation map. 
In the procedure adopted here, this mask depends on the threshold used to select significant signal and the noise of the SAUNAS X-ray surface brightness map. While \citet{pawlik+2016mnras456_3032} use the threshold of $1\sigma$ in their optical images (i.e. the limiting depth of the optical image) to select the galaxy region for asymmetry analysis, they show that the value of $A_S$ decreases when higher thresholds are used (see their Fig. 6). This overall result is reproduced even when the detection mask is defined using the X-ray surface brightness map and is shown in Fig. \ref{fig:ngc5044}. \par  

The segmentation maps used for NGC~5044 are displayed in Fig. \ref{fig:ngc5044} similar to those shown previously in Fig. \ref{fig:masking_connectivity}. Yellow, labeled as 1, show areas of signal and purple, labeled as zero, correspond to areas belonging to the background.  We compare the maps when using 3$\sigma$ and 1.5$\sigma$ X-ray detections in SAUNAS as the threshold (left panels) to derive the final mask $S$ via 8-connectivity (right panels) and resulting $A_S$ values. The figure shows that the $A_S$ value significantly increases when the 1.5$\sigma$ segmentation map is used. This result highlights that the estimation of $A_S$ is highly dependent on the input segmentation map.

To estimate the uncertainty in the $3\sigma$ threshold used for final masking in the measurement of $A_S$, we develop a montecarlo based method by taking advantage of the SAUNAS SNR per pixel map. The SNR map in SAUNAS is derived by dividing the X-ray flux in the SAUNAS flux map per pixel by the 1$\sigma$ dispersion of the flux per pixel. The 1$\sigma$ dispersion per pixel is used to create one thousand flux maps by randomly sampling the flux of each pixel by a gaussian distribution. For each random flux map, the same three step procedure used to create the binary detection mask is followed (see Fig. \ref{fig:masking_connectivity}) and the resulting $A_S$ value measured. This procedure accounts for how the 3$\sigma$ threshold or contour changes due to the noise in the SAUNAS X-ray flux map and consequently, its impact on the binary detection mask and measured shape asymmetry. This procedure is performed for each galaxy in the sample and the uncertainty in $A_S$ is reported as the 1$\sigma$ dispersion of the one thousand measurements made using the random flux maps. The typical uncertainty in $A_S$ derived from this procedure is in the order of 3$\%$.

\begin{figure*}
    \centering
    \includegraphics[width=1.0\linewidth]{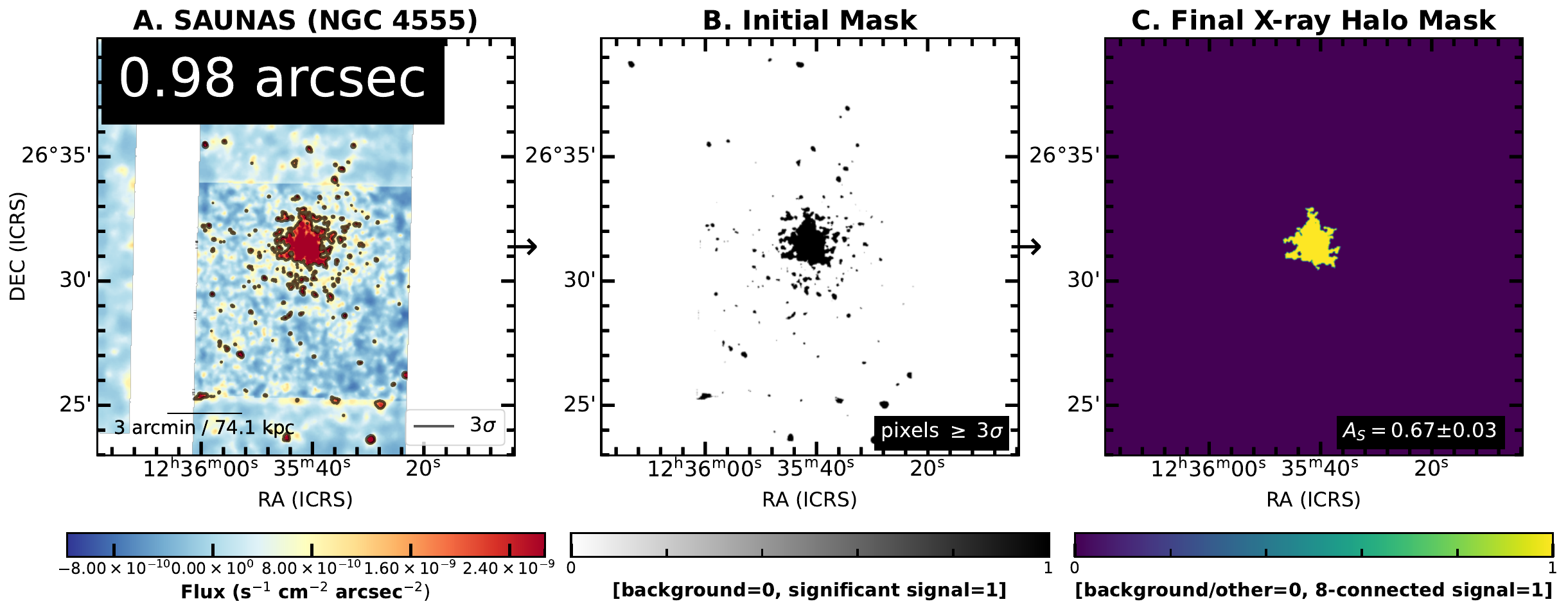}

    \includegraphics[width=1.0\linewidth]{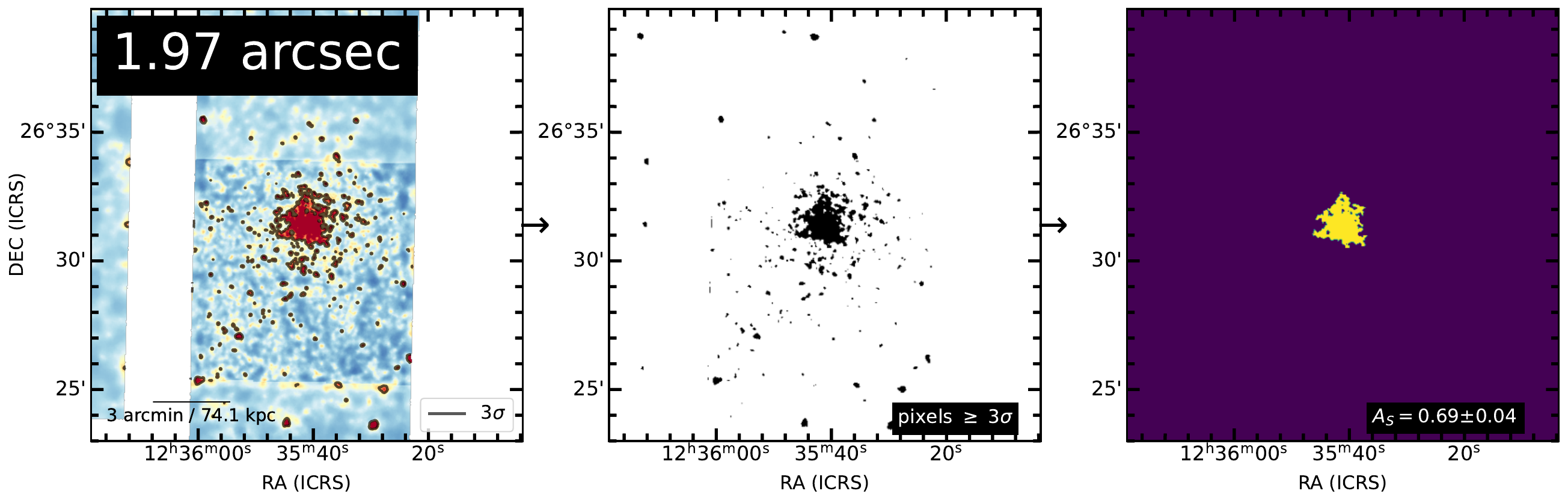}

    \includegraphics[width=1.0\linewidth]{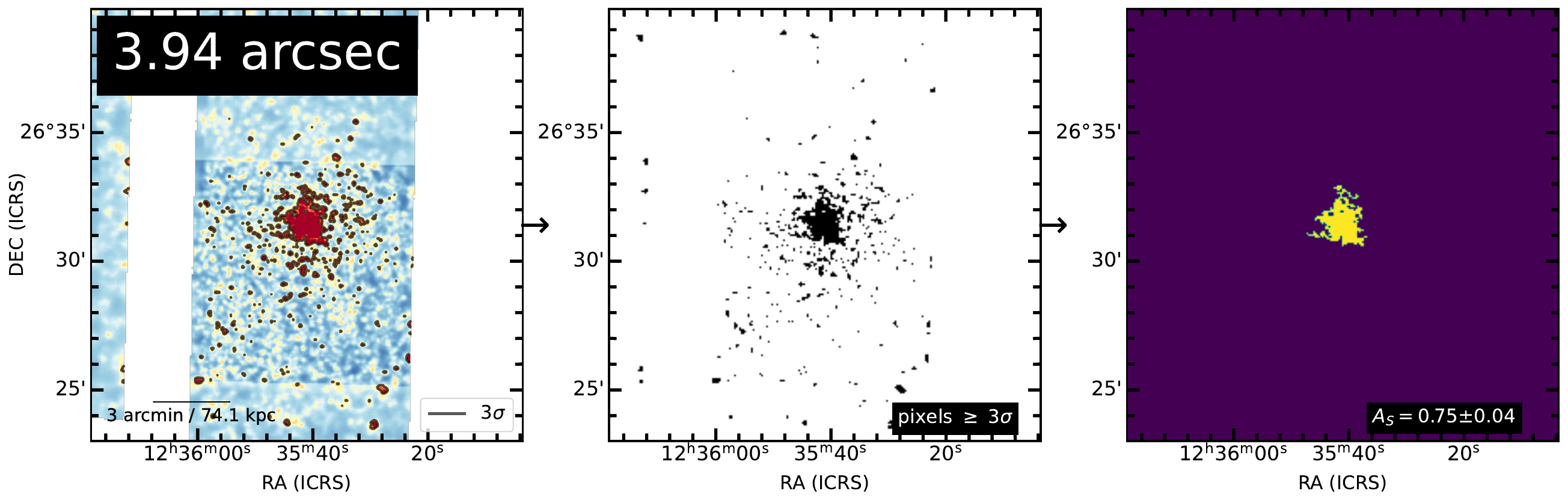}

    \includegraphics[width=1.0\linewidth]{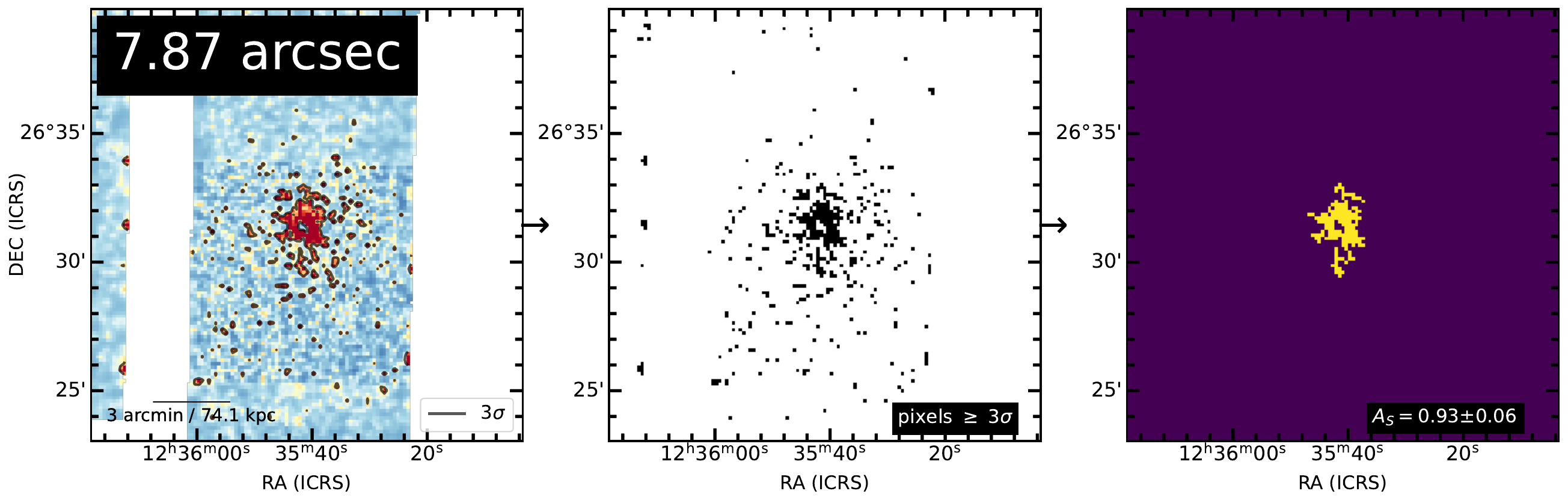}
    
    \caption{Dependence of X-ray halo shape asymmetry on bin size. Panels A-C in each row are as in Fig. \ref{fig:masking_connectivity}. Binary detection masks derived using maps with bin size larger than $\sim$4''significantly increases the value of $A_S$ for galaxy NGC~4555.}
    \label{fig:resolution}
\end{figure*}

Finally, we perform a final test to examine the dependence of the binary detection halo mask on the re-binning of the data. Considering the new X-ray asymmetric halo shape discovered in galaxy NGC~4555, we re-ran the SAUNAS pipeline to obtain its surface brightness maps using bin sizes of 0.98'' and 3.94''. These additional runs give us a total of four maps with increasing bin size. These maps of increasing bin sizes from SAUNAS and their associated binary detection mask are displayed in Fig. \ref{fig:resolution}. \par
The asymmetry values computed using the 0.98'', 1.97'' and 3.94''  maps are compatible within the uncertainty of the measurement. However, the binary detection masks that have a bin size larger than 3.94'' increases the value of $A_S$ for galaxy NGC~4555 significantly. \par

This analysis sets the spatial scale up to which we can characterize the asymmetry in the X-ray halo of NGC~4555 (i.e. up to $\sim 4''$). All the analysis shown in this study used the smaller bin size of 1.97''. At the distance of NGC~4555, $\sim$2'' translate to 0.82\,kpc. For the full sample, 2'' corresponds to 0.57$\pm$0.3\,kpc. Therefore, all the asymmetry analysis presented in this study are applicable to X-ray structures at sub-kpc scales. In a future paper, we plan to examine the impact of bin size and resolution on other physical parameters of X-ray halos.

\section{SAUNAS/\emph{Chandra} maps of X-halos}
\label{app:categories}

\begin{table}[h]
    \centering
    \begin{tabular}{cccccccccc}
    \hline 
Name & RA & DE & Obs. ID & Instrument & Exposure time & Mode & Count rate & Obs. date & PI \\

(1) & (2) & (3) & (4) & (5) & (6) & (7) & (8) & (9) & (10) \\ 
 & [deg] & [deg] & & & [$s$] & & [GHz] & & \\\hline \\

NGC~0193   & 9.83   & 3.33   & 4053  & ACIS-S & 29.06  & VFAINT & 8.91  & 2003-09-01 & O'Dea      \\
                &         &         & 11389 & ACIS-S & 93.92  & VFAINT & 13.64  & 2009-08-21 & Jones      \\
                &         &         &       &        &        &        &                    &                     &            \\
NGC 0383   & 16.85  & 32.41  & 3555  & ACIS-S & 5.08   & FAINT  & 10.45 & 2003-08-06 & Predehl    \\
                &         &         & 2147  & ACIS-S & 44.41  & FAINT  & 10.20 & 2000-11-06 & Hardcastle \\
                &         &         &       &        &        &        &                    &                     &            \\
NGC~0533    & 21.38  & 1.76   & 2880  & ACIS-S & 37.60  & VFAINT & 7.96  & 2002-07-28 & Sarazin    \\
                &         &         &       &        &        &        &                    &                     &            \\
NGC~0741    & 29.09  & 5.63   & 2223  & ACIS-S & 30.35  & FAINT  & 9.67  & 2001-01-28 & Vrtilek    \\
                &         &         & 17198 & ACIS-S & 91.41  & VFAINT & 9.40  & 2015-12-04 & Vrtilek    \\
                &         &         & 18718 & ACIS-S & 58.67  & VFAINT & 9.29 & 2015-12-06  & Vrtilek    \\
                &         &         &       &        &        &        &                    &                     &            \\
IC~1860    & 42.39  & -31.19 & 10537 & ACIS-S & 37.28  & VFAINT & 13.51 & 2009-09-12  & Allen      \\
                &         &         &       &        &        &        &                    &                     &            \\
NGC~1132   & 43.21  & -1.28  & 801   & ACIS-S & 13.99  & FAINT  & 7.76  & 1999-12-10 & Zabludoff  \\
                &         &         & 3576  & ACIS-S & 39.66  & FAINT  & 10.87 & 2003-11-16 & Garmire    \\
                &         &         &       &        &        &        &                    &                     &            \\

NGC~1395 & 54.62 & -23.03 & 799 & ACIS-I & 27.37 & FAINT & 8.25 & 1999-12-31 & Jones \\ 
                &         &         &       &        &        &        &                    &                     &            \\
NGC~1399   & 54.62  & -35.45 & 4172  & ACIS-I & 44.50  & VFAINT & 4.96 & 2003-05-26  & Scharf     \\
                &         &         & 9530  & ACIS-S & 59.35  & VFAINT & 15.79  & 2008-06-08  & Irwin      \\
                &         &         & 26675 & ACIS-S & 20.38  & FAINT  & 8.68  & 2023-08-02  & Dage       \\
                &         &         & 27748 & ACIS-S & 19.71  & FAINT  & 9.43  & 2023-07-30  & Dage       \\
                &         &         & 14527 & ACIS-S & 27.79  & FAINT  & 11.06 & 2013-07-01  & Irwin      \\
                &         &         & 14529 & ACIS-S & 31.62  & FAINT  & 10.98 & 2015-11-06  & Irwin      \\
                &         &         & 16639 & ACIS-S & 29.67  & FAINT  & 11.03 & 2014-10-12  & Irwin      \\
                &         &         & 240   & ACIS-S & 43.53  & FAINT  & 10.07 & 2000-06-16  & Canizares  \\
                &         &         & 49898 & ACIS-S & 13.08  & FAINT  & 9.80   & 2000-06-15  & Canizares  \\
                &         &         & 319   & ACIS-S & 56.04  & FAINT  & 11.52  & 2000-01-18  & Mushotzky  \\
                &         &         & 239   & ACIS-I & 3.60   & FAINT  & 7.24  & 2000-01-19  & Mushotzky  \\
                &         &         & 320   & ACIS-I & 3.38   & VFAINT & 152.29 & 1999-10-18  & Mushotzky  \\
                &         &         & 4174  & ACIS-I & 45.67  & VFAINT & 5.26  & 2003-05-28  & Scharf     \\
                &         &         & 9798  & ACIS-S & 18.30  & VFAINT & 12.34 & 2007-12-24 & Roelofs    \\
                &         &         & 9799  & ACIS-S & 21.29  & VFAINT & 12.47  & 2007-12-27  & Roelofs    \\
                &         &         &       &        &        &        &                    &                     &            \\
NGC~1407        & 55.05  & -18.58 & 7849  & ACIS-S & 4.89   & FAINT  & 16.43 & 2007-07-11  & Mathur     \\
 &         &         & 14033 & ACIS-S & 54.35  & VFAINT & 8.57  & 2012-06-17  & Su         \\
      &         &         & 791   & ACIS-S & 48.57  & VFAINT & 9.72  & 2000-08-16  & White      \\
                &         &         &       &        &        &        &                    &                     &            \\

NGC~1550        & 64.91  & 2.41   & 5800  & ACIS-S & 44.54  & VFAINT & 12.53 & 2005-10-22  & dupke      \\
                &         &         & 3186  & ACIS-I & 9.99   & VFAINT & 4.88  & 2002-01-08  & Murray     \\
                &         &         & 3187  & ACIS-I & 9.65   & VFAINT & 4.87  & 2002-01-08  & Murray     \\
                &         &         & 5801  & ACIS-S & 44.45  & VFAINT & 12.36 & 2005-10-24  & dupke      \\
                &         &         &       &        &        &        &                    &                     &            \\

       \hline 
          
    \end{tabular}
    \caption{\emph{Chandra} archival datasets analyzed in this work. \emph{Chandra}/ACIS observations available within 17~arcmins of each galaxy, retrieved from the \emph{Chandra} Data Archive, as of March 2025. Col (1) Galaxy name; Col.(2) Right Ascension (ICRS); Col. (3) Declination (ICRS); These are the central coordinates of the galaxy used in the analysis. Col (4) Observation ID; Col.(5) instrument or configuration; Col.(6) total exposure time per observation; Col.(7) observing mode, filter or band; Col.(8) average count rate, wavelength or frequency range; Col.(9) exposure start date. Col. (10) Principal Investigator.}
    \label{tab:chandra_archive_datasets}
\end{table}

\begin{table}[]
\centering
\begin{tabular}{cccccccccc}

NGC~1600        & 67.92  & -5.09  & 21374 & ACIS-S & 25.72  & VFAINT & 11.50   & 2018-12-03  & Walker     \\
                &         &         & 21375 & ACIS-S & 42.21  & VFAINT & 11.86 & 2019-11-28  & Walker     \\
                &         &         & 21998 & ACIS-S & 13.87  & VFAINT & 11.51 & 2018-12-03  & Walker     \\
                &         &         & 22878 & ACIS-S & 44.97  & VFAINT & 11.88  & 2019-11-25  & Walker     \\
                &         &         & 22911 & ACIS-S & 31.01  & VFAINT & 11.71 & 2019-11-01  & Walker     \\
                &         &         & 22912 & ACIS-S & 35.64  & VFAINT & 11.66 & 2019-11-02  & Walker     \\
                &         &         & 4283  & ACIS-S & 26.78  & VFAINT & 8.51  & 2002-09-18  & Sarazin    \\
                &         &         & 4371  & ACIS-S & 26.75  & VFAINT & 7.79  & 2002-09-20 & Sarazin    \\
                &         &         &       &        &        &        &                    &                     &            \\
                
NGC~2563        & 125.15 & 21.07  & 7925  & ACIS-I & 48.11  & VFAINT & 3.74 & 2007-09-18 & Mulchaey   \\
                &         &         &       &        &        &        &                    &                     &            \\
NGC~4104         & 181.66 & 28.17  & 6939  & ACIS-S & 35.88  & VFAINT & 10.36 & 2006-02-16 & Buote      \\
                &         &         & 5914  & ACIS-I & 2.15   & VFAINT & 3.14 & 2005-02-27 & Murray     \\
                &         &         &       &        &        &        &                    &                     &            \\
NGC~4261        & 184.85 & 5.83   & 9569  & ACIS-S & 100.94 & FAINT  & 16.26 & 2008-02-12 & Zezas      \\
                &         &         & 834   & ACIS-S & 34.40  & VFAINT & 5.73   & 2000-05-06 & Birkinshaw \\
                &         &         &       &        &        &        &                    &                     &            \\
NGC~4325        & 185.78 & 10.62  & 3232  & ACIS-S & 30.08  & VFAINT & 6.15  & 2003-02-04  & Ponman     \\
                &         &         &       &        &        &        &                    &                     &            \\
NGC~4374 (M84)  & 186.27 & 12.89  & 803   & ACIS-S & 28.47  & VFAINT & 9.06  & 2000-05-19 & Finogenov  \\
                &         &         & 5908  & ACIS-S & 46.08  & VFAINT & 10.84 & 2005-05-01 & Jones      \\
                &         &         & 6131  & ACIS-S & 40.93  & VFAINT & 12.13 & 2005-11-07 & Jones      \\
                &         &         & 20539 & ACIS-S & 39.54  & VFAINT & 12.14 & 2019-04-05 & Russell    \\
                &         &         & 20540 & ACIS-S & 30.17  & VFAINT & 12.01 & 2019-02-26 & Russell    \\
                &         &         & 20541 & ACIS-S & 11.29  & VFAINT & 11.79 & 2019-04-10  & Russell    \\
                &         &         & 20542 & ACIS-S & 34.61  & VFAINT & 12.00 & 2019-03-18 & Russell    \\
                &         &         & 20543 & ACIS-S & 54.34  & VFAINT & 11.78 & 2019-04-27 & Russell    \\
                &         &         & 21845 & ACIS-S & 27.70  & VFAINT & 11.91 & 2019-03-28  & Russell    \\
                &         &         & 21852 & ACIS-S & 15.60  & VFAINT & 14.16 & 2019-02-18 & Russell    \\
                &         &         & 21867 & ACIS-S & 23.63  & VFAINT & 13.06 & 2019-03-13 & Russell    \\
                &         &         & 22113 & ACIS-S & 21.82  & VFAINT & 12.10 & 2019-02-20 & Russell    \\
                &         &         & 22126 & ACIS-S & 35.10  & VFAINT & 11.98 & 2019-02-28 & Russell    \\
                &         &         & 22127 & ACIS-S & 22.77  & VFAINT & 11.87 & 2019-03-02 & Russell    \\
                &         &         & 22128 & ACIS-S & 23.75  & VFAINT & 11.88 & 2019-03-03 & Russell    \\
                &         &         & 22142 & ACIS-S & 20.77  & VFAINT & 12.96 & 2019-03-14 & Russell    \\
                &         &         & 22143 & ACIS-S & 22.75  & VFAINT & 12.79 & 2019-03-16 & Russell    \\
                &         &         & 22144 & ACIS-S & 31.75  & VFAINT & 11.85 & 2019-03-15 & Russell    \\
                &         &         & 22153 & ACIS-S & 21.08  & VFAINT & 13.19 & 2019-03-23 & Russell    \\
                &         &         & 22163 & ACIS-S & 35.59  & VFAINT & 11.84 & 2019-03-29 & Russell    \\
                &         &         & 22164 & ACIS-S & 32.63  & VFAINT & 11.80 & 2019-03-31 & Russell    \\
                &         &         & 22166 & ACIS-S & 38.56  & VFAINT & 12.03 & 2019-04-06  & Russell    \\
                &         &         & 22174 & ACIS-S & 49.41  & VFAINT & 11.81 & 2019-04-11  & Russell    \\
                &         &         & 22175 & ACIS-S & 27.20  & VFAINT & 11.79 & 2019-04-12 & Russell    \\
                &         &         & 22176 & ACIS-S & 51.39  & VFAINT & 11.87 & 2019-04-13 & Russell    \\
                &         &         & 22177 & ACIS-S & 36.58  & VFAINT & 11.85 & 2019-04-14 & Russell    \\
                &         &         & 22195 & ACIS-S & 38.07  & VFAINT & 11.78 & 2019-04-28 & Russell    \\
                &         &         & 22196 & ACIS-S & 20.58  & VFAINT & 12.79 & 2019-05-07 & Russell    \\
                &         &         & 401   & ACIS-S & 1.67   & FAINT  & 6.67  & 2000-04-20 & Garmire    \\
                &         &         &       &        &        &        &                    &                     &            \\
NGC 4555        & 188.92 & 26.52  & 2884  & ACIS-S & 29.97  & FAINT  & 10.59 & 2003-02-04  & Ponman     \\
                &         &         &       &        &        &        &                    &                     &            \\
 
    \end{tabular}
    \caption{Continued from Table \ref{tab:chandra_archive_datasets}}
    \label{tab:chandra_archive_datasets_2}
\end{table}

\begin{table}[]
    \centering
    \begin{tabular}{cccccccccc}

 
NGC 4636        & 190.71 & 2.69   & 323   & ACIS-S & 52.38  & FAINT  & 10.69 & 2000-01-26 & Mushotzky  \\
                &         &         & 324   & ACIS-I & 8.38   & VFAINT & 36.22  & 1999-12-04 & Mushotzky  \\
                &         &         & 3926  & ACIS-I & 74.70  & VFAINT & 4.11  & 2003-02-14  & Jones      \\
                &         &         & 4415  & ACIS-I & 74.36  & VFAINT & 4.09  & 2003-02-15  & Jones      \\
                &         &         &       &        &        &        &                    &                     &            \\
NGC 4782        & 193.65 & -12.56 & 3220  & ACIS-S & 49.33  & VFAINT & 7.31  & 2002-06-16 & Sakelliou  \\
                &         &         &       &        &        &        &                    &                     &            \\
NGC 5044        & 198.85 & -16.39 & 798   & ACIS-S & 20.46  & FAINT  & 18.42  & 2000-03-19  & Goudfrooij \\
                &         &         & 9399  & ACIS-S & 82.68  & VFAINT & 16.48  & 2008-03-07  & David      \\
                &         &         & 17195 & ACIS-S & 78.03  & VFAINT & 12.38 & 2015-06-06 & David      \\
                &         &         & 17196 & ACIS-S & 88.87  & VFAINT & 11.80 & 2015-05-11 & David      \\
                &         &         & 17653 & ACIS-S & 35.53  & VFAINT & 11.51 & 2015-05-07 & David      \\
                &         &         & 17654 & ACIS-S & 25.04  & VFAINT & 11.81  & 2015-05-10  & David      \\
                &         &         & 17666 & ACIS-S & 88.55  & VFAINT & 12.78 & 2015-08-23  & David      \\
                &         &         & 3225  & ACIS-S & 83.14  & VFAINT & 11.18 & 2002-06-07 & KAHN       \\
                &         &         & 3664  & ACIS-S & 61.33  & VFAINT & 11.34 & 2002-06-06 & KAHN       \\
                &         &         &       &        &        &        &                    &                     &            \\

NGC~5129 & 201.04 & 13.98 & 6944 & ACIS-S & 20.91 & VFAINT & 10.58 & 2006-04-13 & Buote\\
& & & 7325 & ACIS-S & 25.84 & VFAINT & 10.61 & 2006-05-14 & Buote\\ 
                &         &         &       &        &        &        &                    &                     &            \\

NGC~5846         & 226.62 & 1.61   & 7923  & ACIS-I & 90.02  & VFAINT & 4.28  & 2007-06-12 & Forman     \\
                &         &         & 788   & ACIS-S & 29.86  & FAINT  & 24.09 & 2000-05-24 & Trinchieri \\
                &         &         &       &        &        &        &                    &                     &            \\
IC~1262          & 263.26 & 43.76  & 6949  & ACIS-I & 38.60  & VFAINT & 4.34 & 2006-04-17 & Forman     \\
                &         &         & 7321  & ACIS-I & 37.55  & VFAINT & 4.380  & 2006-04-19  & Forman     \\
                &         &         & 7322  & ACIS-I & 37.54  & VFAINT & 4.34  & 2006-04-22 & Forman     \\
                &         &         & 2018  & ACIS-S & 30.74  & FAINT  & 12.47 & 2001-08-23 & Trinchieri \\
                &         &         &       &        &        &        &                    &                     &            \\
NGC~6861         & 301.83 & -48.37 & 3190  & ACIS-I & 22.81  & VFAINT & 3.30 & 2002-07-26 & Murray     \\
                &         &         & 11752 & ACIS-I & 93.49  & VFAINT & 4.26  & 2009-08-13 & Machacek   \\
                &         &         &       &        &        &        &                    &                     &            \\
NGC~7619        & 350.06 & 8.21   & 3955  & ACIS-S & 37.47  & VFAINT & 8.50   & 2003-09-24  & Kim        \\
                &         &         & 2074  & ACIS-I & 26.74  & VFAINT & 4.24  & 2001-08-20 & Forman  \\ \\\hline  
    \end{tabular}
    \caption{Continued from Table \ref{tab:chandra_archive_datasets_2}.}
    \label{tab:chandra_archive_datasets_3}
\end{table}

\begin{figure}[h!]
    \centering
    \includegraphics[width=0.32\linewidth]{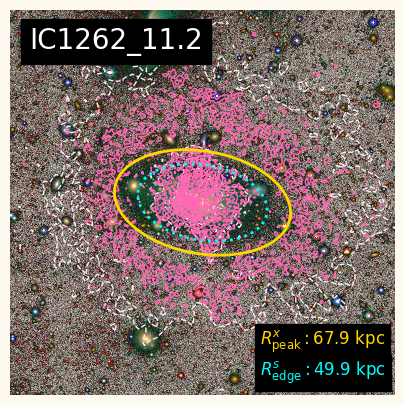} 
    \includegraphics[width=0.32\linewidth]{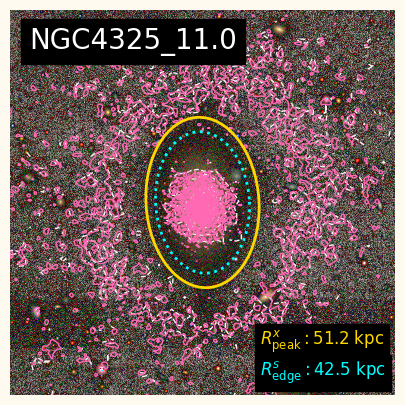}
    \includegraphics[width=0.32\linewidth]{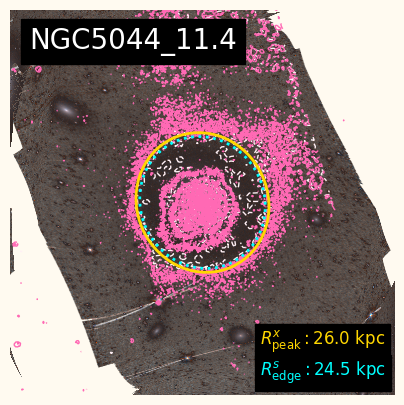}
    \caption{Galaxies where $A_S < 0.4$ do not have significant asymmetries in their X-ray halo. The case of NGC~5044 is discussed in more detail in the text and Fig. \ref{fig:ngc5044}.}
    \label{fig:no_asym}
\end{figure}

\begin{figure}[h!]
    \centering
    \includegraphics[width=0.32\linewidth]{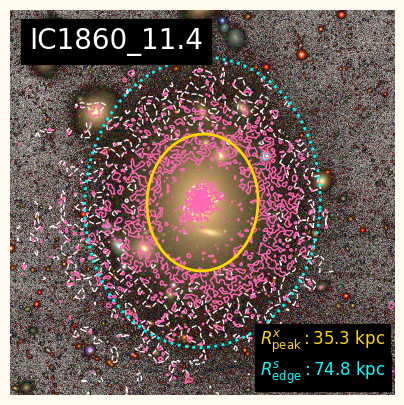}
    \includegraphics[width=0.32\linewidth]{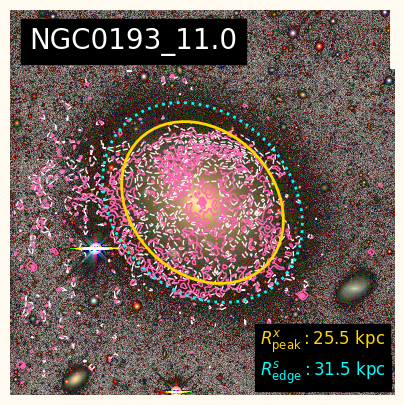}  
    \includegraphics[width=0.32\linewidth]{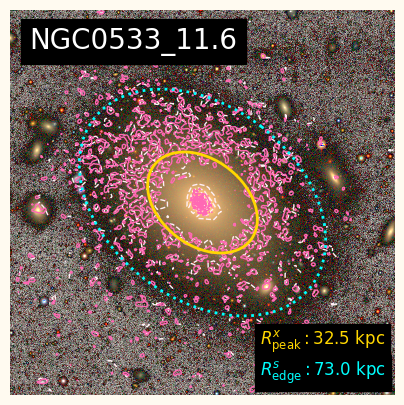}
    \includegraphics[width=0.32\linewidth]{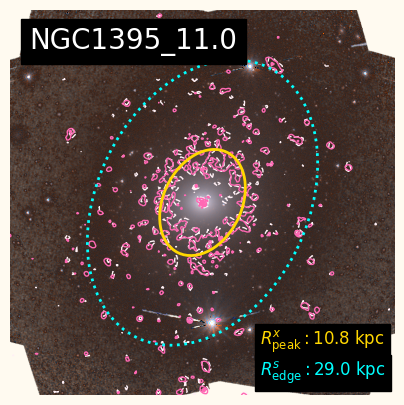}
    \includegraphics[width=0.32\linewidth]{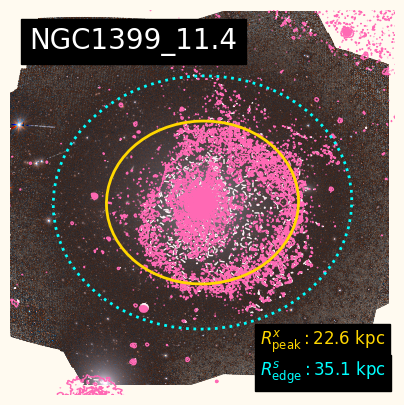}
    \includegraphics[width=0.32\linewidth]{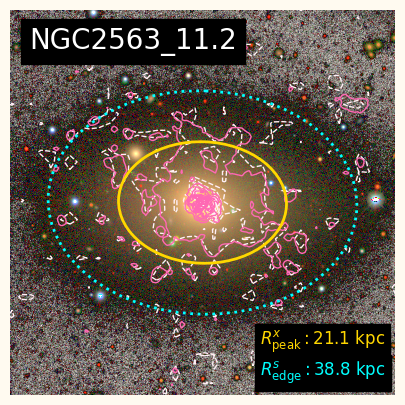}
    \includegraphics[width=0.32\linewidth]{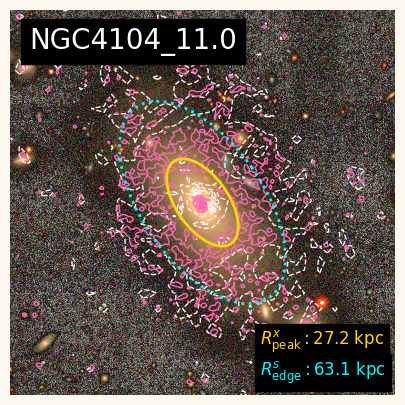}
    \includegraphics[width=0.32\linewidth]{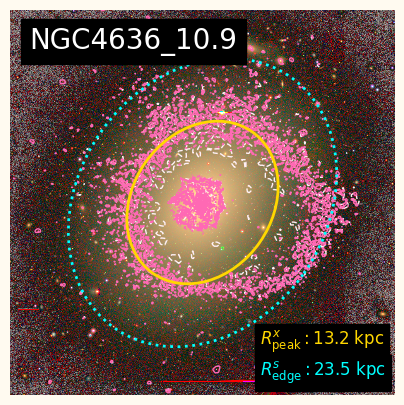}
    \includegraphics[width=0.32\linewidth]{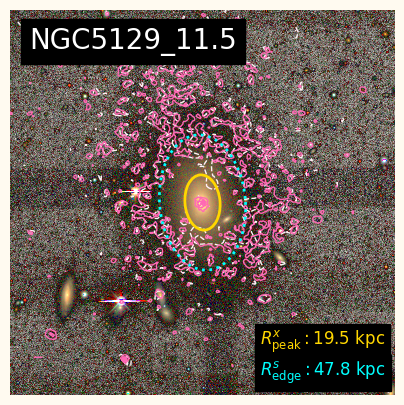}
    \includegraphics[width=0.32\linewidth]{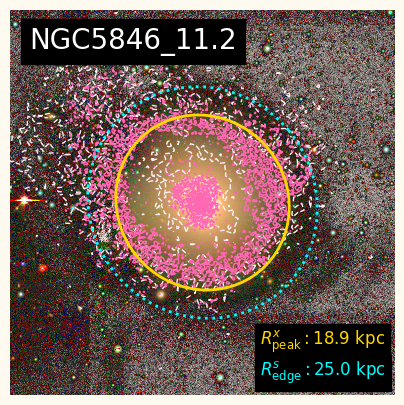}
    \includegraphics[width=0.32\linewidth]{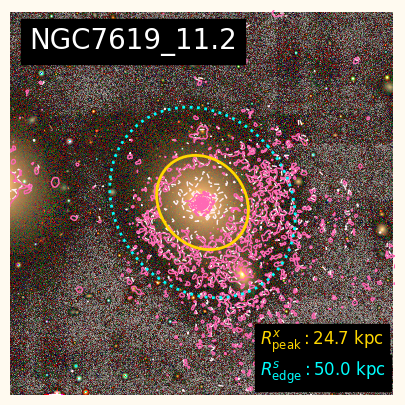}
    \caption{Galaxies with asymmetry $A_S > 0.4$}
    \label{fig:def_asym}
\end{figure}

SAUNAS maps for the rest of the \rpeak{} sample  which were not shown as part of the main text are presented in Fig. \ref{fig:no_asym}--\ref{fig:def_asym}. The archival datasets used as input for SAUNAS for the full sample are listed in Table \ref{tab:chandra_archive_datasets}-\ref{tab:chandra_archive_datasets_3}.

\bibliography{pruned.bib}{}
\bibliographystyle{aasjournal}

\end{document}